\documentclass[a4paper]{article}
\usepackage{newlfont}
\usepackage{amsfonts}
\usepackage{amsmath}
\usepackage{graphicx}
\usepackage{mathrsfs}
\usepackage{MnSymbol}
\usepackage{setspace}
\usepackage{cases}
\usepackage{stmaryrd}
\usepackage{empheq}
\usepackage{scrextend}

\newcommand{\g}{\gamma}
\newcommand{\Q}{\mathbf{q}}
\newcommand{\Rot}{\mathbf{R}}
\newcommand{\V}{\mathbf{V}}
\newcommand{\B}{\mathbf{B}}
\newcommand{\A}{\mathbf{A}}
\newcommand{\C}{\mathbf{C}}

\newcommand{\Og}{\mathcal{O}}
\newcommand{\p}{\varphi}
\newcommand{\R}{\mathbf{r}}

\newcommand{\E}{\mathbf{e}}

\newcommand{\M}{\mathbf{m}}
\newcommand{\f}{\mathbf{f}}

\newcommand{\pa}{\! \! \Arrownot \, \, \Arrownot \, \, \, \,}

\begin{document}
\title{\textbf{Motion planning and motility maps for flagellar microswimmers} }
\author{Giancarlo Cicconofri\thanks{Corresponding author.} \, and Antonio DeSimone \\
SISSA, International School for Advanced Studies
\\
 Via Bonomea 265, 34136 Trieste - Italy \\
 giancarlo.cicconofri$@$sissa.it, desimone$@$sissa.it}

\date{}
\maketitle

\begin{abstract}
We study two microswimmers consisting of a spherical rigid head and a passive elastic tail. In the first one the tail is clamped to the head, and the system oscillates under the action of an external torque. In the second one, head and tail are connected by a joint allowing the angle between them to vary periodically, as a result of an oscillating internal torque. Previous studies on these models were restricted to sinusoidal actuations, showing that the swimmers can propel while moving on average along a straight line, in the direction given by the symmetry axis around which beating takes place. We extend these results to motions produced by generic (non-sinusoidal) periodic actuations within the regime of small compliance of the tail. We find that modulation in the velocity of actuation can provide a mechanism to select different directions of motion. With velocity modulated inputs the externally actuated swimmer can translate laterally with respect to the symmetry axis of beating, while the internally actuated one is able to move along curved trajectories. The governing equations are analysed with an asymptotic perturbation scheme, providing explicit formulas whose results are expressed through motility maps. Asymptotic approximations are further validated by numerical simulations.
\end{abstract}

\thispagestyle{empty}

\newpage

\tableofcontents

\thispagestyle{empty}

\newpage

\section{Introduction}

Flagella constitute the means of propulsion for a large variety of swimming microorganisms and for bio-inspired robots targeted to medical applications \cite{Stone,Lauga_Swimmer}. In eukaryotes these long and flexible appendages are typically actuated by distributed internal forces. Mammalian spermatozoa, for example, propagate bending waves along their tails to achieve propulsion \cite{Hermes1}. Others, like the bi-flagellate \emph{Chlamydomonas}, perform a rhythmical breaststroke-like routine leading to a rocketing forward motion \cite{Guasto}. 

On the other hand, passive elastic flagella, when actuated only at one extremity, can also constitute a simple but effective swimming device. A biological example is the bacterium  \emph{E. Coli}, whose passive helical tail is actuated at one end by a rotary motor inducing a cork-skew like propulsion \cite{Guasto}. Locomotion at very small scales is subject to the so called ``Scallop Theorem'' \cite{Purcell}, which states that the body of a swimmer must undergo time-irreversible shape changes to produce net advancement. Interestingly, the hydroelastic coupling between a passive elastic filament and the surrounding  fluid constitutes, by itself, a source of time-irreversibility.

Much has been done on this topic. Besides the pioneering work by Machin \cite{Machin}, the problem has been explored extensively in more recent years. In \cite{Gol1,Gol2} Wiggins \emph{et al.} demonstrated that, apart from axial rotations, also the planar beating of an elastic filament with one oscillating end can produce axial propulsive force. These findings have been put in the swimming context by Lauga in \cite{Lauga}, who analysed the locomotion capabilities of an internally actuated swimmer as the one in Figure \ref{fig1}(\textbf{b}). Numerical experiments also focused on externally actuated swimming of microrobots, inspired by the geometry of sperm cells, consisting of a cargo with a clamped passive elastica \cite{Abbott,Hermes2} like the one depicted in Figure \ref{fig1}(\textbf{a}). Both externally and internally actuated elastic swimmers were also analysed through discrete models by Or \emph{et al.} in \cite{Or1,Or2} to grasp the essentials of their motility mechanism. However, in all the aforementioned studies flagellar beating is always restricted to sinusoidal actuations; the swimmers move ``head-first'' and, on average, on a straight line. 

Many questions remain unanswered. For instance, what is the direction of motion for a generic (non-sinusoidal) periodic actuation? Does swimming always take place head-first, or can the sign of swimming velocity be controlled? In this paper we provide an answer to these questions. Our analysis is based on a small-compliance assumption as in \cite{Maxey}, which leads to a simplification of the governing equations and allows for analytic asymptotic approximation. Within this limit we provide explicit formulas whose results are expressed through  motility maps \cite{ADS1,ADS2}, a visual approach that yields the displacement produced by a given actuation without the need to actually solve the dynamics on a case-by-case basis. The results of the asymptotic analysis are further validated by the comparison with numerical simulations.

Our main finding is the following: modulation in the velocity of actuation can provide a mechanism to select different directions of motion, for both model swimmers. In particular, a flagellar oscillation composed by a fast down-beat and slow up-beat produces generally a deviation from the symmetry axis around which beating takes place. With these oscillatory inputs, the externally actuated swimmer of Figure \ref{fig1}(\textbf{a}) can translate laterally with respect to this symmetry axis, while the internally actuated one of Figure \ref{fig1}(\textbf{b}) is able to move along curved trajectories. Moreover, for the externally actuated swimmer, we find a sign reversal  in the average velocity for large enough actuation amplitudes. 

Dependence on the velocity of actuation is not surprising since previous investigations on this kind of model swimmers \cite{Lauga,Or1,Or2} reported a non-linear dependence between displacements and frequency of oscillation. With our analysis we can look deeper into the relations between i) given (generic) actuation, ii) shape changes of the swimmers, and iii) displacement after one actuation cycle. In fact, we demonstrate how the actuation velocity can be considered as a motion control parameter. 

Take first the internally actuated swimmer. In order to obtain a net displacement, the swimmer must undergo non-reciprocal shape changes to overcome the Scallop Theorem.  Here the shape of the swimmer is determined by the angle $\alpha$ between the head and the tail, see Figure \ref{fig1}(\textbf{b}), and by the geometry $y$ of the tail itself. We find that, as a result of the dynamics, $y$ depends at first approximation on only two parameters: the internal angle $\alpha$ and its velocity $\dot{\alpha}$. This last two quantities, then, can be interpreted then as the shape parameters of the swimmer. Non-reciprocal cycles in the shape space $(\dot{\alpha},\alpha)$, result in different displacements and rotations of the swimmer, and the resulting net motions can be inferred with the aid of the motility maps provided in Section \ref{MM_Int}.

A similar analysis is carried out in Section \ref{Ext} for the swimmer of Figure \ref{fig1}(\textbf{a}). Because of the presence of an external torque, in this case the Scallop Theorem does not apply, see \cite{Or2,ADS3,Choset1}.  Non-reciprocity is, on the other hand, still crucial.  Interestingly, in this case the tail geometry is, at first approximation, a standing wave (i.e., a fixed function of space multiplied by a time-dependent amplitude) and it is completely determined by only one parameter: the velocity $\dot{\phi}$. Net displacements arise, as a result of external activation, when the angle $\phi$, see Figure \ref{fig1}(\textbf{a}), and the the geometry of the tail (which is determined by $\dot{\phi}$) undergo a non-reciprocal cycle. Modulating loops in the space $(\dot{\phi},\phi)$ leads to different resulting displacements that can be inferred from the motility maps of Section \ref{MM_Ext}. 

The paper is organized as follows. In Section \ref{Gov} we set  up the governing equations for both models, following closely the derivation in \cite{Lauga}, and we formalize the assumptions on the dynamical parameters. In Section \ref{Ext}  we derive an explicit asymptotic solution for the externally actuated swimmer of Figure \ref{fig1}(\textbf{a}), we derive and comment the motility maps, and we confront the asymptotic solutions with numerical simulations. A similar analysis is carried out in Section \ref{Int} for the internally actuated swimmer of Figure \ref{fig1}(\textbf{b}).

\section{Two model microswimmers} \label{Gov}
\begin{figure}[htbp]
	\centering
		\includegraphics[width=1.00\textwidth]{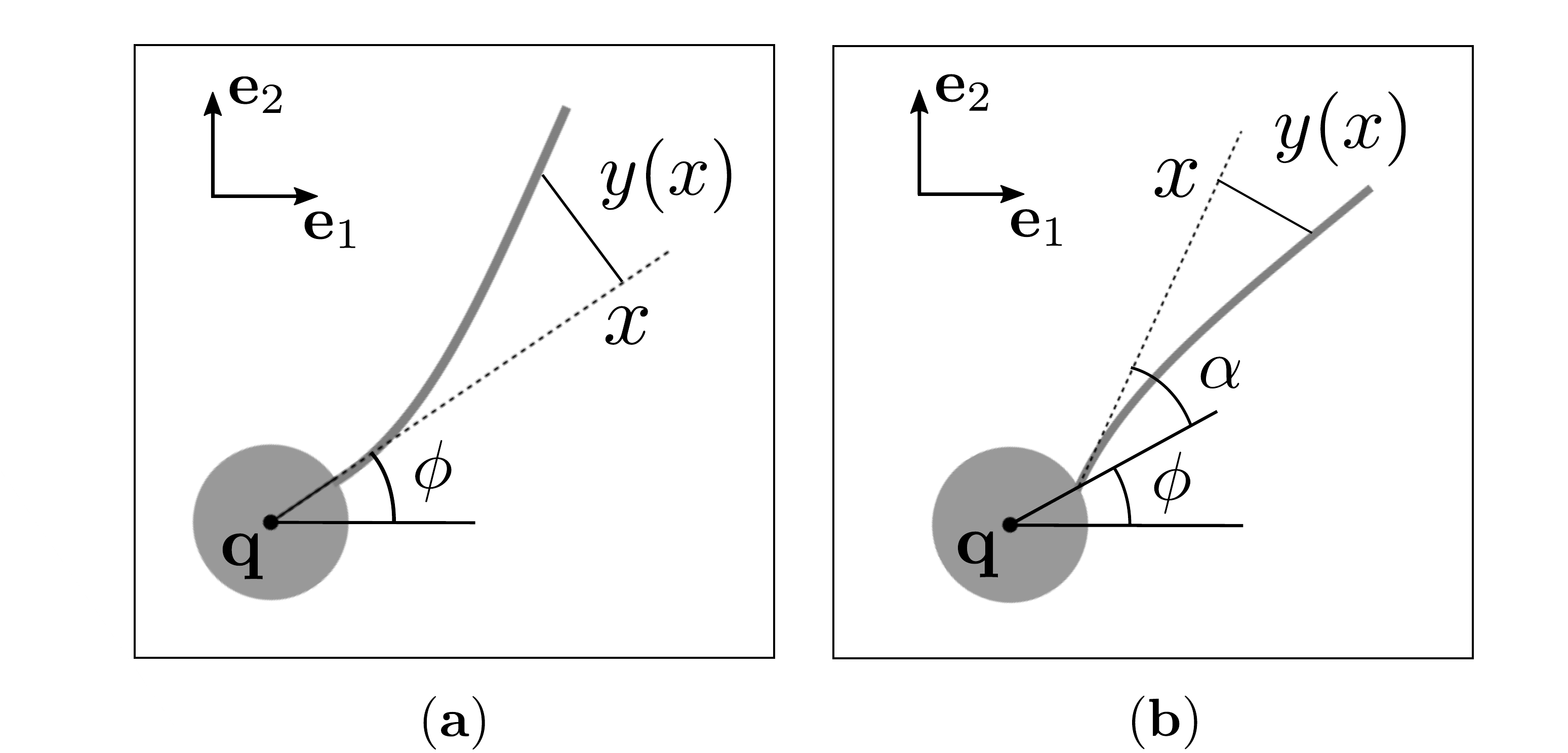}
		\caption{Schematic description of (\textbf{a}) the externally actuated and (\textbf{b}) the internally actuated swimmer model.}
	\label{fig1}
\end{figure}

We consider planar motions of the model swimmers illustrated in Figures \ref{fig1}(\textbf{a}) and \ref{fig1}(\textbf{b}) (although the analysis we present here can be carried out also for fully three dimensional motions) and we suppose that the plane of locomotion is spanned by two orthonormal vectors $\left\{\E_1,\E_2\right\}$. Both models consist of a spherical cargo of radius $a$ attached to a passive elastic filament of length $L$,  the ``flagellum''. Both swimmers are surrounded by a Newtonian fluid, moving at low Reynolds number. 

In the case of Figure \ref{fig1}(\textbf{a}) the swimmer's flagellum is clamped orthogonally to the cargo, and propulsion is achieved due to an external torque acting on the sphere. We suppose that such a torque modulates the angle $\phi$ formed by the horizontal line and the line joining the centre of the cargo and the point of attachment. We define the angle $\phi$ for the swimmer of Figure \ref{fig1}(\textbf{b}) in the same way. Here, however, we suppose that the flagellum is connected to the sphere by a joint, and that the relative angle $\alpha$ between the flagellum and the cargo at the point of attachment can vary, under the action of an internal torque.

In both cases the elastic flagellum is supposed to be inextensible and slender, that is 
\begin{equation}
\frac{r_{f}}{L} \ll 1 \, , \label{slender}
\end{equation}
where $r_{f}$ the radius of its cross-section. As in \cite{Gol1,Gol2,Lauga} we model the dynamics of interaction between the fluid and such a slender filament using ``resistive force theory''. The relationship between forces and velocities is local, and the viscous drag coefficients $\xi_{\pa}$ and $\xi_{\bot}$ are, respectively, the force exerted by the fluid per unit length of the flagellum for motion parallel and perpendicular to its length. For classical reviews on resistive force theory see \emph{e.g.} \cite{Light,BreWin}.

We neglect here hydrodynamic interactions between  flagellum and  cargo. The interaction between the fluid and the cargo is given by classical Stokes drag formulas. The mass of the swimmers is also neglected as viscous and elastic forces dominate the dynamics.

Our analysis relies on yet another hypothesis, previously stated in the introduction: we assume that the flagellum (in both cases) has a large bending resistance compared with the viscous forces applied to it. More precisely, if $B$ is the bending stiffness of the flagellum and $\omega$ the frequency of actuation (either internal or external), then we assume that the ratio between the typical normal viscous force $\xi_{\bot} L^{2} \omega$ and the typical elastic force $B/L^2$ acting on the flagellum 
\begin{equation}
	\epsilon = \frac{\xi_{\bot} \omega }{B} L^{4}
\label{Machin}
\end{equation}
is small. As a consequence, the flagellum does not deviate much from a straight line. In view of this, we take as resistive drag coefficients the ones calculated for straight slender bodies in \cite{Cox}, that is
\begin{equation}
	\xi_{\pa} = \frac{2 \pi \mu}{\log(L/r_{f})+ \lambda_{\pa}} \quad \textrm{and} \quad   \xi_{\bot} = \frac{4 \pi \mu}{\log(L/r_{f})+ \lambda_{\bot}} \, ,\label{xi}
\end{equation}
where $\mu$ is the dynamic viscosity of the fluid, while $\lambda_{\pa}$ and $\lambda_{\bot}$ are constant of the order $\Og(1)$ depending on the cross-sectional shape of the flagellum. In the following we shall refer to $\epsilon$ as the ``Machin number'' and always assume
\begin{equation}
	\epsilon \ll 1 \, . \label{epsilon_fla}
\end{equation} 
Assumption \eqref{epsilon_fla} is satisfied, for example, for a magnetic Permalloy flagellum with $L/r_{f}\simeq 10^3$ and Young modulus $E\simeq 10^{11} \textrm{Nm}^{-2}$, as in \cite{ADS2}. Indeed, if we consider such a flagellum beating in water $\mu = 8.90 \times 10^{ - 4}\textrm{Nm}^{-2}\textrm{s}$, using the formula $B=\pi E r_{f}^4 /4$ we have
\begin{displaymath}
	\epsilon= \frac{8 \mu \omega (L/r_{f})^4}{\big( \log(L/r_{f}) + \lambda_{\bot} \big) E} \:  \lesssim \: \omega \times 10^{-2}\textrm{s}  \, ,
\end{displaymath}
which satisfies \eqref{epsilon_fla} for a reasonable range of frequencies. We point out that this is not the case for all flagellar swimmers of interest: for the artificial swimmer in \cite{Stone} $\epsilon \sim 10^0-10^3$, while for sperm cells $\epsilon \sim 10^2-10^5$ \cite{Hermes1}.

\subsection{Governing equations and simplifying assumptions}
The equations of motion for our systems have been derived already in previous investigations, see \emph{e.g.} \cite{Lauga}. To write them in non-dimensional form we scale the space variables  by $L$, forces by $B/L^2$, moments by $B/L$, and time $t$ by $\omega^{-1}$. The balance of viscous and elastic forces on the flagellum, for both the externally and the internally actuated case, gives 
\begin{equation}
    \epsilon \left(  \textbf{Id} +  (\gamma - 1) \frac{\partial \R}{\partial s} \frac{\partial \R}{\partial s} \right) \cdot \frac{\partial \R}{\partial t}  = - \frac{\partial^4 \R}{\partial s^4} + \frac{\partial}{\partial s}\left( \sigma \frac{\partial \R}{\partial s} \right)    \label{flag_nl} 
\end{equation}
where $\R=\R(s,t)$ is the arc-length parametrized curve (on the plane) describing the position of the flagellum, and $\g=\xi_{\pa}/\xi_{\bot} \sim  0.5$ is the viscous anisotropy ratio. Equation \eqref{flag_nl} is accompanied by that for the Lagrange multiplier $\sigma=\sigma(s,t)$ imposing the inextensibility condition $|\partial \R /\partial s|=1$, see \cite{Lauga}. We omit here this equation since $\sigma$ will drop out from the equations governing the dynamics thanks to the approximations we make in the following.

The (normalized) elastic force $\f_{\text{el}}$ and the elastic moment $\M_{el}$ are given by
\begin{equation}
\f_{\text{el}}= - \frac{\partial^3 \R}{\partial s^3} +  \sigma \frac{\partial \R}{\partial s} \quad \text{and} \quad \M_{\text{el}}= \frac{\partial \R}{\partial s} \times \frac{\partial^2 \R}{\partial s^2} = K(s,t) \E_{3} \, ,
\label{fm_nl}
\end{equation}
where $\E_{3}$ is the unit vector normal to the plane of locomotion and $K$ is the curvature of $\R$. At the free end ($s=1$) we assume that no forces or torques are acting on the flagellum, thus $\f_{\text{el}} (1,t) =\mathbf{0}$ and $\M_{\text{el}} (1,t) =\mathbf{0}$. The force balance for the cargo, for both the externally and the internally actuated case, reads  
\begin{equation}
	- \epsilon  \eta \dot{\Q} (t)  + \f_{\text{el}} (0,t) =\mathbf{0} \, ,\label{fbal_nl}
\end{equation}
where $\Q$ is the (normalized) coordinate of the centre point of the cargo and 
\begin{equation*}
 \eta = \frac{6 \pi \mu a}{\xi_{\bot} \omega L^2} = \frac{3}{2} \rho \big( \log(L/r_{f}) + \lambda_{\bot} \big)
\label{eta}
\end{equation*}
where $\rho=a/L$ is the normalized radius of the cargo. For the externally actuated swimmer the balance of moments (with respect to $\Q$) for the cargo gives
\begin{equation*}
- \epsilon \eta_{\textrm{rot}} \dot{\phi}(t ) \E_{3}  +  (\R(0,t) - \Q(t)) \times \epsilon \eta \dot{\Q}(t) +  \M_{\text{el}}(0,t) +  \boldsymbol \tau_{\text{ext}}(t) = 0 \, , \
\end{equation*}
where $\boldsymbol \tau_{\text{ext}}$ is the (normalized) external torque acting on the cargo and
\begin{equation*}
 \eta_{\textrm{rot}} = \frac{8 \pi \mu a^{3}}{\xi_{\bot} \omega L^3} = 2 \rho^3 \big( \log(L/r_{f}) + \lambda_{\bot} \big) .
\end{equation*}
For the internally actuated swimmer the moment balance reads
\begin{gather*}
- \epsilon \eta_{\textrm{rot}} \dot{\phi}(t ) \E_{3}  + (\R(0,t) - \Q(t)) \times \epsilon \eta \dot{\Q}(t) - \boldsymbol \tau_{\textrm{int}}(t) = \mathbf{0} \quad \textrm{with} \quad \boldsymbol \tau_{\textrm{int}}(t) = -\M_{\text{el}}(0,t) \, ,
\end{gather*}
where $\boldsymbol \tau_{\textrm{int}}$ is the (normalized) internal torque acting on the flagellum modulating the angle difference $\alpha$. 

As mentioned in the previous section, we restrict our analysis to the case in which the flagellum bends mildly away from a straight moving reference axis (the dotted line in Figures  \ref{fig1}(\textbf{a}) and \ref{fig1}(\textbf{b}) ). We assume that the direction of the reference axis is given at time $t$ by the normal vector
\begin{displaymath}
	\E_{\theta} = \cos \theta(t) \, \E_{1} + \sin \theta(t) \E_{2} \, ,
\end{displaymath}
where $\theta=\phi$ in the externally actuated case and $\theta=\phi+\alpha$ in the internally actuated one. Following \cite{Lauga}, we represent the curve $\R$ as
\begin{equation}
	\R  =   \R(0,t) + x \E_{\theta} + y(x,t) \E_{\theta}^{\bot} \, , \label{r}
\end{equation}
where $\E_{\theta}^{\bot}= - \sin \theta(t) \, \E_{1} + \cos \theta(t) \E_{2}$ is the unit vector orthogonal to the reference axis. We suppose 
\begin{equation}
y (0,t) =0   \quad \textrm{and} \quad \frac{\partial y}{\partial x} (0,t) =0
\label{attach}
\end{equation}
so that, in particular, the orientation of the flagellum at the point of attachment is determined by $\theta$. Moreover, we assume the scalar function $y$ to be small, along with all its derivatives. In fact, as we can validate a-posteriori, $y$ and its derivatives can be assumed to be of order $\Og(\epsilon)$. In this regime the variable $x$ can be considered as the arc-length coordinate of the curve $\R$, instead of $s$. Using the approximation $\partial/\partial s \simeq \partial/\partial x$ we obtain
\begin{equation}
\frac{\partial \R}{\partial s} \simeq  \E_{\theta} + \frac{\partial y}{\partial x} \E_{\theta}^{\bot} \: \: , \quad \frac{\partial^2 \R}{\partial s^2} \simeq  \frac{\partial^2 y}{\partial x^2} \E_{\theta}^{\bot}\: \: , \quad \text{and} \quad \: \: \frac{\partial^3 \R}{\partial s^3} \simeq  \frac{\partial^3 y}{\partial x^3} \E_{\theta}^{\bot}
\label{ders}
\end{equation}
therefore we can write
\begin{equation}
	\f_{\text{el}} =  \sigma \E_{\theta}  + \left( - \frac{\partial^3 y}{\partial x^3} + \sigma \frac{\partial y}{\partial x} \right)\E_{\theta}^{\bot} \quad \textrm{and} \quad \M_{\text{el}}= \frac{\partial^2 y}{\partial x^2} \E_{3} \, . \label{fm}
\end{equation}
The boundary conditions at the free edge ($x=1$) read then
\begin{equation}
	\sigma(1,t)=0 \: \: ,\quad  \frac{\partial^2 y}{\partial x^2} (1,t) =0 \: \: ,  \quad \textrm{and} \quad \frac{\partial^3 y}{\partial x^3} (1,t) =0 \label{BC}
\end{equation}
while, projecting \eqref{fbal_nl} on $\E_{\theta}$ and $\E_{\theta}^{\bot}$, we have
\begin{equation}
	- \epsilon  \eta \dot{\Q}(t) \cdot \E_{\theta}  + \sigma (0,t) = 0 \quad \textrm{and} \quad 	- \epsilon  \eta \dot{\Q}(t) \cdot \E_{\theta}^{\bot}  -  \frac{\partial^3 y}{\partial x^3} (0,t)  = 0 \, .
\label{fbal}
\end{equation}
The moment balance for the cargo for the externally actuated swimmer becomes
\begin{equation}
 - \epsilon \eta_{\textrm{rot}} \dot{\phi}(t ) \E_{3}  + (\R(0,t)-\Q(t)) \times \epsilon \eta \dot{\Q}(t) + \frac{\partial^2 y}{\partial x^2} (0,t)\E_{3} +  \boldsymbol \tau_{\text{ext}}(t) = 0  \label{mbal_ext}
\end{equation}
while for the internally actuated swimmer we have
\begin{gather}
 - \epsilon \eta_{\textrm{rot}} \dot{\phi}(t ) \E_{3} + (\R(0,t)-\Q(t)) \times \epsilon \eta \dot{\Q}(t) - \boldsymbol \tau_{\textrm{int}}(t) = \mathbf{0} \label{mbal_int} \\ \quad \textrm{with} \quad \boldsymbol \tau_{\textrm{int}}(t) = - \frac{\partial^2 y}{\partial x^2}(0,t) \E_{3}  \, . \nonumber
\end{gather}
Finally, we rewrite the force balance equations on the flagellum \eqref{flag_nl} with $\R$ given by \eqref{r}. Notice first that
\begin{equation}
	\frac{\partial \R}{\partial t} = \Big(u^x - \dot{\theta}y\Big) \E_{\theta} + \Big(u^y + x\dot{\theta} + \frac{\partial y}{\partial t}\Big)\E_{\theta}^{\bot} \, , \label{dert}
\end{equation}
where
\begin{equation}
u^x= \frac{\partial \R}{\partial t}(0,t) \cdot \E_{\theta} \quad \textrm{and} \quad u^y= \frac{\partial \R}{\partial t}(0,t) \cdot \E_{\theta}^{\bot} \, .
\label{uxuy}
\end{equation}
Using \eqref{ders} and \eqref{dert} we express the left hand side of equation \eqref{flag_nl} in terms of $u^x$, $u^y$, $\theta$, and $y$. We make here a simplification: we drop terms of this expression involving powers of $y$ and its derivatives, as they are negligible within our small bending assumption. Projecting on $\E_{\theta}$ and $\E_{\theta}^{\bot}$ we obtain
\begin{empheq}[left=\empheqlbrace]{align}
& \: \epsilon \left(\g \left( u^x - \dot{\theta}y \right) +(\g -1) \left( u^y + x\dot{\theta} \right) \frac{\partial y}{\partial x}\right)  = \frac{\partial \sigma}{\partial x} \label{flag1} \\
& \: \epsilon \left( u^y + x\dot{\theta} + \frac{\partial y}{\partial t} +(\g - 1)u^{x}\frac{\partial y}{\partial x}\right)  =  - \frac{\partial^4 y}{\partial x^4} +  \frac{\partial}{\partial x}\left(\sigma \frac{\partial y}{\partial x}\right) \label{flag2}
\end{empheq}
In Sections \ref{Ext} and \ref{Int}, further simplifications of these equations are derived. 

We end this section with a comment. The simplification scheme we adopted here follows very closely the one proposed in \cite{Lauga}. In \cite{Lauga}, however, small bendings of the flagellum come from the hypothesis of small actuation  amplitude, while here they arise as a consequence of small compliance \eqref{epsilon_fla}. While the orientation of the axis of reference (the dotted line in Figures \ref{fig1}(\textbf{a}) and \ref{fig1}(\textbf{b}) ) of the flagellum is considered fixed in \cite{Lauga}, here we allow large amplitude oscillations of this axis. Moreover, we drive our system with periodic yet generic inputs, while only sinusoidal actuations are considered in \cite{Lauga}.

\section{Externally actuated swimmer} \label{Ext}
The natural problem for the externally actuated swimmer is that of finding the motion given the external actuation torque $\boldsymbol \tau_{\textrm{ext}}$. We study here first the problem in which $\phi$ is given rather than $\boldsymbol \tau_{\textrm{ext}}$, for two reasons: first, it simplifies the asymptotic calculations; second, the mechanism generating propulsion is better understood when treated in terms of the configurational parameter $\phi$. Observe that, in this case, the moment balance equation \eqref{mbal_ext} is only used a-posteriori, to determine the external torque $\tau_{\textrm{ext}}$ needed to impose the prescribed oscillations of $\phi$. 

We first simplify further the system of equations. As mentioned in the previous section, we can assume that $y$ and its derivatives are of order $\Og(\epsilon)$. In addition, since $\theta=\phi$, from \eqref{BC}, \eqref{fbal}, and \eqref{flag1} we have
\begin{displaymath}
	0=-\epsilon \eta \dot{\Q} \cdot \E_{\theta} + \sigma(0,t) = -\epsilon \eta u^x - \int_{0}^{1} \frac{\partial \sigma}{\partial x} \, dx = -\epsilon(\eta + \gamma) u^x + \Og(\epsilon^2)
\end{displaymath}  
which implies $u^x =\Og( \epsilon)$, and therefore $\sigma = \Og( \epsilon^2)$. The terms multiplied by $u^x$ and $\sigma$ in \eqref{flag2} are of order $\Og(\epsilon^3)$, and can be dropped from the equation. The transversal force balance on the flagellum then reads
\begin{equation}
 \epsilon \left(u^{y}(t) + \dot{\phi}(t)x + \frac{\partial y}{\partial t}(x,t)\right)  = - \frac{\partial^4 y}{\partial x^4}(x,t) \, . \label{ext1}
\end{equation}
Notice that the variables $y$ and $\sigma$ are now decoupled. Integrating $\sigma$ from \eqref{flag1} and using definitions \eqref{uxuy}, equations \eqref{fbal} can be written as
\begin{empheq}[left=\empheqlbrace]{align}
& \: - \epsilon(\eta +\g )u^x + \epsilon \g  \int_{0}^{1} \dot{\phi} y \,dx  - \epsilon (\g-1) \int_{0}^{1}(u^y + x \dot{\phi})\frac{\partial y}{\partial x} \, dx = 0 \label{ext2} \\
& \: - \epsilon \eta u^{y}(t) + \epsilon \eta \rho \dot{\phi}(t)  - \frac{\partial^3 y}{\partial x^3}(0,t) = 0 \label{ext3}
\end{empheq}
With the boundary conditions \eqref{attach} and \eqref{BC} we can solve \eqref{ext1}-\eqref{ext3} for the unknowns $u^{x}$, $u^{y}$, and $y$ once we prescribe the initial condition $y(0,x)$ for the deviation of the flagellum. The system is solved numerically with a finite difference scheme, based on the one proposed in \cite{She}. Here, and in all the numerical simulations presented in the paper, we take $r_{f}/L= 10^{-3}$, $\lambda_{\pa}=-0.5$, and $\lambda_{\bot}=0.5$.

In the next section we derive a perturbation scheme to obtain a formal asymptotic solution for the system of equations. In Section \ref{MM_Ext} we compare numerical solutions and analytical approximations, and we discuss physical interpretations of the results.

\subsection{Asymptotics}\label{Asym_Ext}
We now proceed formally in finding an asymptotic solution of our problem, applying standard perturbation techniques. We look for solutions in the form of a power series in the Machin number
\begin{equation}
	y = y_{0} + \epsilon y_{1} + \epsilon^{2} y_{2} + \ldots \: \: , \quad u^{x} = u^{x}_{0} + \epsilon u^{x}_{1} + \epsilon^{2} u^{x}_{2} + \ldots \: \: , \quad u^{y} = u^{y}_{0} + \epsilon u^{y}_{1} + \epsilon^{2} u^{y}_{2} + \ldots  \label{asym}
\end{equation}
For $\phi$ we take $\phi=\phi_0 + \epsilon \phi_1  + \epsilon^2 \phi_2 \ldots$, where $\phi_{0}=\phi$ and $\phi_{k}=0$ for $k\geq 1$. Substituting these expression in \eqref{ext1}-\eqref{ext3} and expanding all members of the equations in power series of $\epsilon$, we equate coefficients of like powers of $\epsilon$. We obtain a series of equations to be solved successively. Equation \eqref{ext1} becomes
\begin{equation}
0 = - \frac{\partial^4 y_{0}}{\partial x^4}(x,t) \quad \textrm{and} \quad  u^{y}_{k-1}(t) + \dot{\phi}_{k-1}(t)x + \frac{\partial y_{k-1}}{\partial t}(x,t)  = - \frac{\partial^4 y_{k}}{\partial x^4}(x,t) \label{ext1k}
\end{equation} 
for $k\geq 1$, while \eqref{ext2} gives
\begin{equation}
0 = - \frac{\partial^3 y_{0}}{\partial x^3}(0,t) \quad \textrm{and} \quad  -\eta u_{k-1}^{y}(t) + \eta \rho \dot{\phi}_{k-1}(t) = \frac{\partial^3 y_{k}}{\partial x^3}(0,t)  \label{ext2k}
\end{equation}
for $k\geq 1$. These equations come with the following boundary conditions
\begin{equation}
y_k(0,t)= \frac{\partial y_k}{\partial x} (0,t)= \frac{\partial^2 y_k}{\partial x^2} (1,t)=\frac{\partial^3 y_k}{\partial x^3} (1,t)=0 \quad \textrm{for $k\geq 0$} \, . \label{BCk}
\end{equation}
Notice that $u^x$ is completely decoupled from $u^y$ and $y$, as it appears only in \eqref{ext3}. We can then solve (formally) for $u^y$ and $y$ from  \eqref{ext1k} and \eqref{ext2k}, and subsequently recover the asymptotic expression for $u^x$ though the equality \eqref{ext3}. Observe also that equations \eqref{ext1k} come from the expansion of an  equation \eqref{ext1} in which the only derivative with respect to time is multiplied by $\epsilon$. As a consequence, we can not impose the initial condition $y(0,x)$ on the asymptotic solution of $y$. This is a well known aspect of this kind of perturbation schemes, in which solutions approximate the unperturbed ones apart from an initial ``boundary layer''.

In the following we calculate explicitly the asymptotic solution up to order $k=1$. Clearly, at order zero we must have
\begin{displaymath}
	y_{0}(x,t)=0 \, .
\end{displaymath}
Equation \eqref{ext1k} for $k=1$ then reads
\begin{equation}
	u^{y}_{0}(t) + \dot{\phi}(t)x = - \frac{\partial^4 y_{1}}{\partial x^4}(x,t) \, . \label{eq_y1}
\end{equation}
The unique solution for $y_1$ satisfying the previous equation and the boundary conditions \eqref{BCk} can be written as
\begin{equation}
	y_{1}(x,t) = - \int_{0}^{x} \! \int_{0}^{x_1} \! \! \int_{x_2}^{1} \! \int_{x_3}^{1} \Big( u^{y}_{0}(t) + \dot{\phi}(t)x_4 \Big) \,  dx_{4}  dx_{3}  dx_{2}  dx_{1} \, . \label{y1}
\end{equation}
Substituting \eqref{y1} in \eqref{ext2k} we have
\begin{displaymath}
	\eta \rho \dot{\phi}(t) -  \eta u_{0}^{y}(t) = \int_{0}^{1} \Big( u^{y}_{0}(t) + \dot{\phi}(t)x \Big) \,  dx 
\end{displaymath}
which gives
\begin{equation}
	 u_{0}^{y}(t) = \frac{\eta \rho - \frac{1}{2}}{\eta +1} \, \dot{\phi}(t) \, . \label{uy0}
\end{equation}
Plugging the above expression for $u_0^y$ back in \eqref{y1} we obtain 
\begin{equation}
\begin{split}
& \quad \quad  \quad \quad  y_{1}(x,t) =  -p_1(x) \dot{\phi}(t) \, , \quad \textrm{where}  \\
& p_1(x)=  \int_{0}^{x} \! \int_{0}^{x_1} \! \! \int_{x_2}^{1} \! \int_{x_3}^{1} \left( \frac{\eta \rho - \frac{1}{2}}{\eta +1} + x_4  \right)\,  dx_{4}  dx_{3}  dx_{2}  dx_{1} \, . 
\end{split}
\label{p1}
\end{equation}
is a polynomial in $x$ whose coefficients can be calculated explicitly. Notice that, at leading order, the shape of the flagellum $y=\epsilon y_1 + \Og(\epsilon^2)$  is completely determined by the actuation \emph{velocity} $\dot{\phi}(t)$. This  is not entirely surprising, since the bending of the flagellum must be proportional to the total moment applied to it, which itself depends on the velocity of the swimmer. We will return on this observation in the next section.

We now find an explicit solution for $ u_{1}^{y}$ by considering the $k=2$ order problem in \eqref{ext1k}, which reads
\begin{displaymath}
	u_{1}^{y}(t) - p_1(x)\ddot{\phi}(t) =  - \frac{\partial^4 y_{2}}{\partial x^4}(x,t) \, . 
\end{displaymath}
Following the same arguments as in the case of $y_1$ we obtain the integral formula 
\begin{displaymath}
	y_{2}(x,t) = - \int_{0}^{x} \! \int_{0}^{x_1} \! \! \int_{x_2}^{1} \! \int_{x_3}^{1} \Big( u^{y}_{1}(t) - p_1(x_4)\ddot{\phi}(t) \Big) \,  dx_{4}  dx_{3}  dx_{2}  dx_{1}
\end{displaymath}
which, substituted in \eqref{ext2k} for $k=2$, gives
\begin{displaymath}
 -  \eta u_{1}^{y}(t) = \int_{0}^{1} \Big( u^{y}_{1}(t) - p_1(x)\ddot{\phi}(t) \Big) \,  dx \, .
\end{displaymath}
From the above equation we have
\begin{equation}
	u_{1}^{y}(t) = U_{y}^{1}\ddot{\phi}(t)   \quad \text{where} \quad U_{1}^{y}=\frac{\int_{0}^{1}p_1 }{\eta +1}\, . \label{uy1}
\end{equation}
We can now find the solution for the first orders of $u^x$. If we replace the asymptotic expansion of $y$ in \eqref{ext3} we get
\begin{eqnarray*}
	 u^{x} & = & \frac{\g}{\eta+\g} \int_{0}^{1} \dot{\phi}\epsilon  y_1 -  \frac{\g -1}{\eta+\g} \int_{0}^{1} (u_{0}^y + x \dot{\phi})\epsilon \frac{\partial y_1}{\partial x} + \Og(\epsilon^2)  \\
	& = & \epsilon \frac{\g}{\eta+\g} \int_{0}^{1} \dot{\phi} y_1 +  \epsilon \frac{\g -1}{\eta+\g} \int_{0}^{1} \frac{\partial^4 y_1}{\partial x^4}\frac{\partial y_1}{\partial x} + \Og(\epsilon^2)  \\
	& = & \epsilon \frac{\g}{\eta+\g} \int_{0}^{1} \dot{\phi} y_1 +  \epsilon \frac{\g -1}{2} \left(\frac{\partial^2 y_1}{\partial x^2}\right)^2 \! \! (0,t) + \Og(\epsilon^2)  
\end{eqnarray*}
from which we obtain the expressions for the orders $k=0$ and $k=1$ of the $u^x$ asymptotic expansion  
\begin{equation}
 	u_{0}^{x}=0 \: \: \: \:  \textrm{and} \: \: \: \:  u_{1}^{x} = U_{1}^{x} \dot{\phi}^2 \: \: \: \: \text{where}  \: \: \: \: U_{1}^{x} = - \left(  \frac{\g}{\eta+\g}\int_{0}^{1}p_1 \: + \frac{1-\g}{2(\eta+\g)} \, \frac{d^2 p_{1}}{dx^2}(0)^2 \right) \, . \label{ux}
\end{equation}

\subsection{Swimming trajectories and motility maps}\label{MM_Ext}
We now consider the motion of the swimmer given a (generic) periodic actuation $\phi$. The resulting trajectories of the coordinate $\Q$ of the cargo are obtained by integrating the expression
\begin{equation}
	\dot{\Q}=\frac{d}{dt} \Big( \R(0,t) -\rho \E_{\phi} \Big) =u^x \E_{\phi} + u^y \E_{\phi}^{\bot} - \rho \dot{\phi}\E_{\phi}^{\bot} \, , \label{qdot_ext}
\end{equation}
where $u^{x}$ and $u^y$ are either the numerical or the asymptotic solutions of \eqref{ext1}-\eqref{ext3}. 

Notice that, if $\phi$ is periodic, then the approximated asymptotic solutions for $u^x$ and $u^y$ are also periodic. Numerical solutions of \eqref{ext1}-\eqref{ext3}, while depending on the initial condition $y(0,x)$, converge after very few oscillations ($n\sim 3$ for most simulations) to a periodic orbit which turn out to be independent from $y(0,x)$. Asymptotic solutions provide a good approximation of numerical solutions past this short time transient. This is the time range to which we restrict our analysis in the following. 

An important consequence of the periodicity of solutions is that, since $u^x$ and $u^y$ converge to periodic functions, \eqref{qdot_ext} implies that $\dot{\Q}$ is also periodic. If we consider the net displacement of the cargo after the $n^{\textrm{th}}$ period $\Delta_n \Q$, then 
\begin{displaymath}
	\Delta_{n} \Q = \int_{n}^{n+1} \dot{\Q} = \int_{n'}^{n'+1} \dot{\Q} = \Delta_{n'} \Q =: \Delta \Q
\end{displaymath}
for every $n$ and $n'$. The definition of $\Delta \Q$ is well posed, and it gives the net displacement of the cargo after \emph{any} period of the actuation. The displacement $\Delta \Q$ is the main focus of this section.

 Without loss of generality we can restrict our analysis  to the case in which the  angle $\phi$ oscillates around $0$ (i.e. the flagellum is beating around the horizontal axis). As it is intuitive, an oscillation about a given angle produces the same kinematics of swimming, up to a rotation, as when the same oscillation is performed about the horizontal.   

We discuss first the asymptotic solution of \eqref{qdot_ext}, based on the results we obtained in the previous section. We  expand $\dot{\Q}$ into a power series in the Machin number up to the first order $\dot{\Q} = \dot{\Q}_{0} + \epsilon \dot{\Q}_{1} + \Og(\epsilon^2)$. From \eqref{uy0},  \eqref{uy1}, and \eqref{ux} we have
\begin{equation}
\dot{\Q}_0 =\left(\frac{\eta \rho - \frac{1}{2}}{\eta +1}-\rho \right)\dot{\phi}\E_{\phi}^{\bot} \quad \text{and} \quad \dot{\Q}_1= U_{1}^{x} \dot{\phi}^{2} \E_{\phi} +  U_{1}^{y} \ddot{\phi}\E_{\phi}^{\bot} \, . \label{dq0dq1}
\end{equation}
From this expansion we obtain a formula for the displacement $\Delta \Q = \Delta \Q_0 + \epsilon \Delta \Q_1 + \Og(\epsilon^2)$. At order zero we have
\begin{equation}
	\Delta \Q_0 = \int_{n}^{n+1} \dot{\Q}_0 =\int_{n}^{n+1} \left(\frac{\eta \rho - \frac{1}{2}}{\eta +1}-\rho \right)\frac{d}{dt}\E_{\phi} = \mathbf{0} \, . \label{Dq0}
\end{equation}
Notice that $\Q_0$ can be seen as the trajectory of a swimmer with a rigid straight flagellum, since for $\epsilon=0$  equations \eqref{ext1}-\eqref{ext3} describe precisely this system. Equation \eqref{Dq0} then says that a rigid swimmer, whose periodic orientation is controlled externally, always undergoes reciprocal motions, and it is incapable of net advancements.

Now, the equation for $\dot{\Q}_1$ in \eqref{dq0dq1} has the following form 
\begin{equation}
\dot{\Q}_1=\A(\phi) \dot{\phi}^2 + \B(\phi) \ddot{\phi} \, .
\label{AB}
\end{equation} 
where $\A(\phi)=U_{1}^{x} \E_{\phi}$ and $\B(\phi)=U_{1}^{y}\E_{\phi}^{\bot}$. It can be easily shown that the right hand side of \eqref{AB} gives, in general, a non-zero result when integrated over one period of $\phi$. We do that by deducing an integral formula for $\Delta \Q_{1}$. Indeed, suppose  that the (closed) curve given by $t \to (\dot{\phi}(t),\phi(t)) \subset \mathbb{R}^{2}$ parametrizes the boundary $\partial \Omega$ of a domain $\Omega$ in $\mathbb{R}^{2}$ (not to be confused with the plane of locomotion). Observe that, with the position $\psi=\dot{\phi}$, \eqref{AB} can be rewritten as
\begin{equation}
\dot{\Q}_1=\A(\phi) \psi \dot{\phi} + \B(\phi) \dot{\psi} \, .
\label{connection}
\end{equation}
By Stokes theorem we obtain
\begin{equation}
	 \Delta \Q_{1}  = \int_{\partial \Omega} \A(\phi) \psi d\phi  + \B(\phi) \, d\psi  =  \int_{\Omega} \left(\A(\phi) -\frac{d \B}{d \phi}(\phi)\right) \, d\psi d \phi \, .
 \label{stokes}
\end{equation}
We have $\A-d \B /d\phi=C \E_{\phi}$, where $C=U_{1}^{x}+U_{1}^{y}$. From \eqref{p1}, \eqref{uy1}, and \eqref{ux} we can explicitly calculate
\begin{equation}
	C=-(1 -\g) \left( \frac{5 + 12 \eta (2 + 5 \rho) + 
   4 \eta^2 (7 + 42 \rho + 
      45 \rho^2)}{1440 (1 + \eta )^2 (\g + \eta)}\right) \, .
\label{C}
\end{equation}
Observe that since $ \g \neq 1$, then $C \neq 0$. Thus, the integral in \eqref{stokes} will be non-zero in general.  As in \cite{Lauga}, drag anisotropy  is essential to achieve locomotion. Indeed, from the previous equations, we can write the approximated expression for $\Delta \Q$ as
\begin{equation}
	\Delta \Q  = \epsilon \! \! \int_{\Omega} \V \, d\psi d \phi + \Og(\epsilon^{2})    \label{Dq}
\end{equation}
where $\V(\psi , \phi) = C \left(\cos \phi \E_1 + \sin \phi \E_2 \right)$ is a map from $\mathbb{R}^{2}$ with values in the locomotion plane $\left\{\E_{1}, \E_{2}\right\}$, while  $\Omega$ is the planar domain whose boundary is given by 
\begin{equation}
\partial \Omega=\left\{(\dot{\phi}(t),\phi(t)) \, ; \, t \in \left[0,1\right]\right\} \subset \mathbb{R}^{2} \, . \label{boundary}
\end{equation}
Notice that the right hand side of \eqref{Dq} depends on the given actuation $t \mapsto \phi(t)$ only through the integration set $\Omega$.  We can quantify the direction and magnitude of the displacement $\Delta \Q$ simply by guessing the geometry of $\Omega$ and then, with the visual aid of a plot of the vector field $\V$ (a ``motility map''), by estimating the integral in \eqref{Dq}. In the following we show some examples of this estimating procedure, illustrating motility maps for different actuations.

\subsubsection*{Actuation velocity as a control parameter}

\begin{figure}[h]
	\centering
		\includegraphics[width=1.00\textwidth]{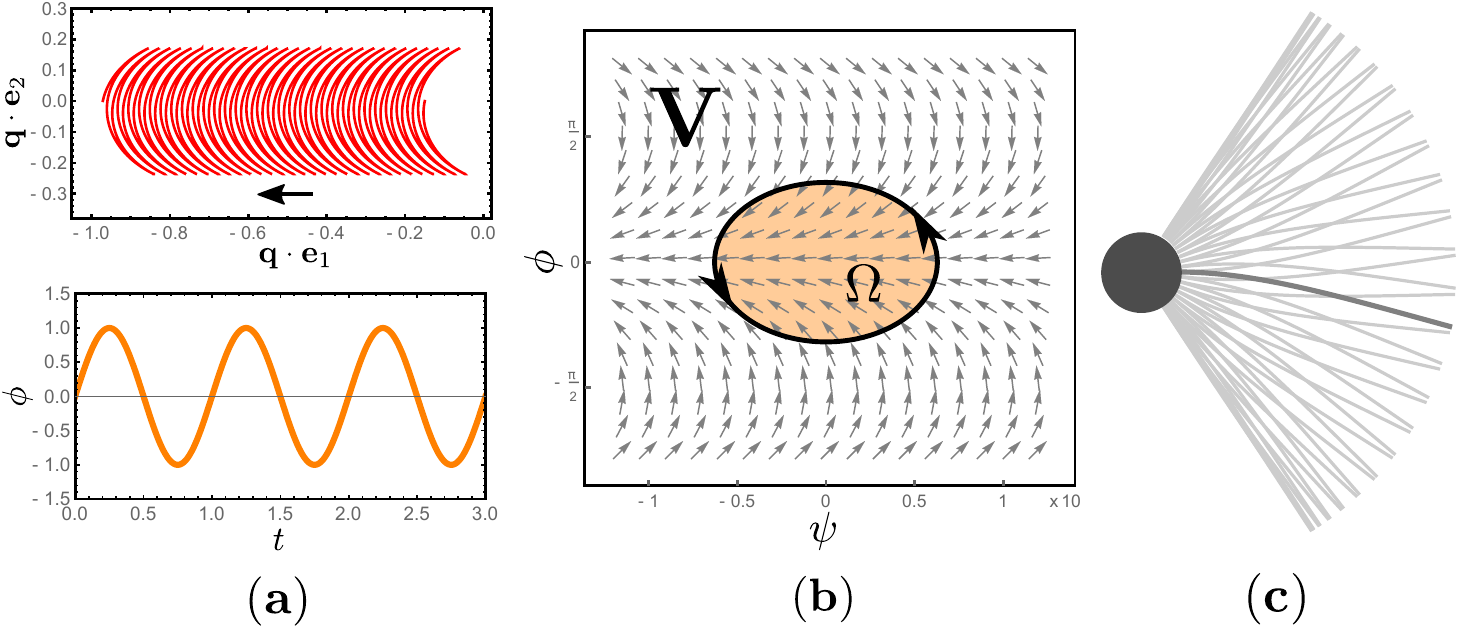} 
	\caption{Sinusoidal oscillations of the system with $\rho=0.15$ and $\epsilon=0.4$. (\textbf{a}) The actuation $\phi$ (bottom) and the numerical solution for $\Q$ (top). (\textbf{b}) Motility map. (\textbf{c}) Snap-shots of the swimmer in motion, as seen by a frame attached to the centre of the cargo.}
	\label{fig2}
\end{figure}

Let us consider first the simplest example, namely that of a sinusoidal actuation $\phi(t)=\sin 2\pi t$, as in Figure \ref{fig2}(\textbf{a}). In this case $\partial \Omega$ is the ellipse  centred in the origin pictured in Figure \ref{fig2}(\textbf{b}). From \eqref{C} we have that $C<0$. As Figure \ref{fig2}(\textbf{b}) shows, the horizontal component $\V \cdot \E_1$ of the vector field $\V$ is negative at every point of the integration set $\Omega$. On the other hand, the projection  $\V\cdot \E_2$ is an odd function of the variable $\phi$. Thus, because of the symmetry of $\Omega$, the $\E_2$ component of the integral in \eqref{Dq} vanishes. Formula \eqref{Dq} predicts then a motion along the horizontal direction, from right to left. This is confirmed by the numerical solution depicted in Figure \ref{fig2}(\textbf{a}), and it is also well known from the literature \cite{Abbott,Hermes2}. 

\begin{figure}[h]
	\centering
		\includegraphics[width=1.00\textwidth]{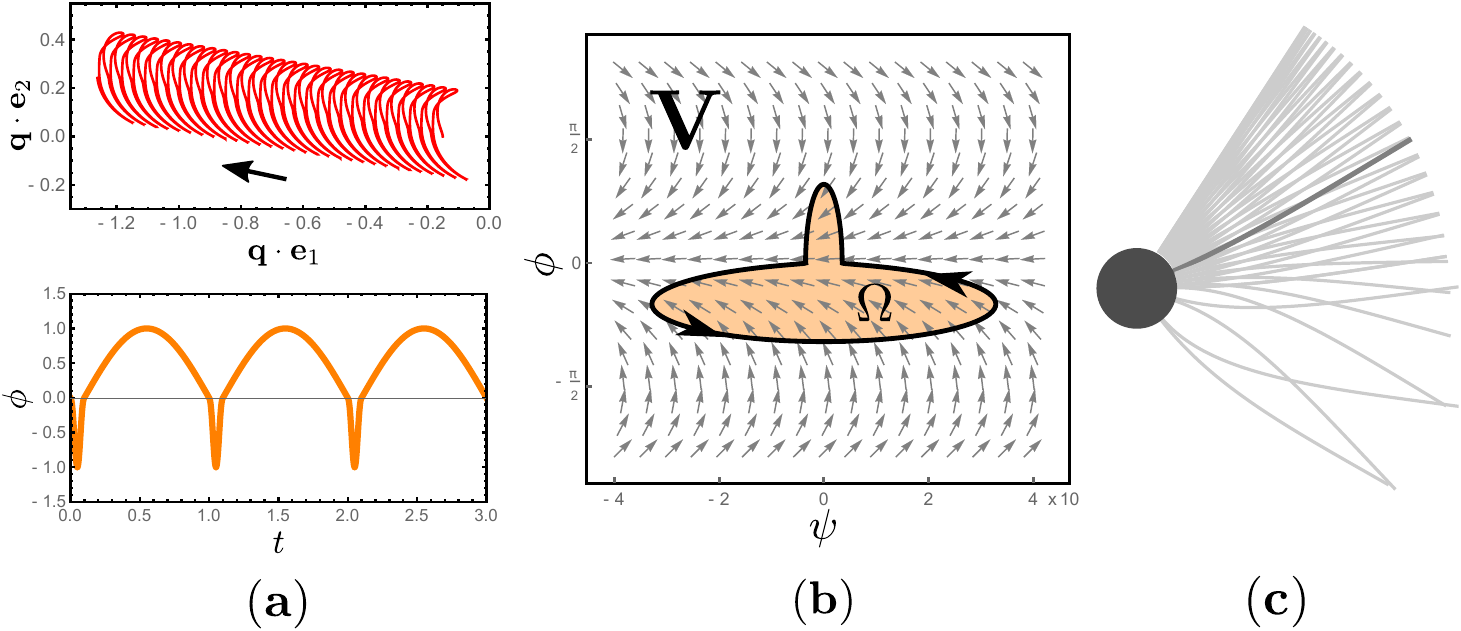}
\caption{Non-sinusoidal oscillations of the system with $\rho=0.15$ and $\epsilon=0.4$. (\textbf{a}) The actuation $\phi$ (bottom) and the numerical solution for $\Q$ (top). (\textbf{b}) Motility map. (\textbf{c}) Snap-shots of the swimmer in motion, as seen by a frame attached to the centre of the cargo.}
	\label{fig3}
\end{figure}

We now take the non-sinusoidal actuation $\phi(t)$ shown in Figure \ref{fig3}(\textbf{a}). The amplitude of the oscillations is the same as before. However, we now have $\dot{\phi}$ small when $\phi>0$, and $\dot{\phi}$ large when $\phi<0$. This leads to a non-symmetric integration set $\Omega$. The set is larger in the region where $\V\cdot \E_2>0$ and, as a result, the swimmer moves with a positive vertical displacement at every cycle. In order to obtain a negative displacement in the vertical direction we can consider the ``negative'' of the previous angle evolution $t \mapsto -\phi(t)$. The resulting set $\Omega$ is the reflection about the $\psi$ axis of the previous one, and the vertical displacement of the swimmer in one cycle changes sign.

Formula \eqref{Dq} generalizes a crucial observation on the motility of externally controlled elastic swimmers made in \cite{Or2}. In this work the authors analyse a two-link swimmer with a passive elastic joint. They conclude that, to achieve propulsion, the orientation of one arm of the swimmer and the internal angle between the links (the ``shape'' of the swimmer) must undergo non-reciprocal cycles, thus breaking the time-reversibility of the interaction with the Stokes fluid. Focusing on sinusoidal oscillations, they demonstrate that the elastic joint ``makes it happen''. In fact, the competition between elastic restoring  forces and viscous drag generates a phase lag in the shape response, and hence non-reciprocal motion.

The exact same mechanism holds for our swimmer, as we can deduce from  formula \eqref{Dq}. As we mentioned in Section \ref{Asym_Ext}, at leading order the bending of the flagellum $y=- \epsilon p_1 \dot{\phi} + \Og(\epsilon^2)$ is fully described by the angle velocity. This is because of the forces bending the flagellum are of viscous nature, hence proportional to $\dot{\phi}$. Formula \eqref{Dq} tells, in particular, that we have a finite displacement,  at least at leading order, only if the measure of the set $\Omega$ is non-zero: that is, only if $\phi$ and $\dot{\phi}$, and therefore the orientation $\phi$ and the ``shape'' $y$, undergo non-reciprocal cycles.  

A sinusoidal actuation leads to $\phi$ and $\dot{\phi}$ out of phase and it is sufficient to produce locomotion along the symmetry axis of beating. However, we show that not only the model \emph{can} swim, but the actuation velocity $\dot{\phi}$ can be as well used as a control parameter to obtain lateral motion with respect to this axis.

We can push this analogy further, by putting our results in the context of geometric control theory \cite{Coron}. Indeed, problems of motion control arising in different fields such as wheeled robot locomotion \cite{KeMu} and crawling \cite{ADS1}  all  lead to systems of equations similar to \eqref{connection}. We can, in fact, rewrite \eqref{connection} in the general form
\begin{equation}
	\left(\begin{array}{c} \dot{\psi} \\ \dot{\phi} \\ \dot{\Q}_1 	\end{array} \right) = \left(\begin{array}{c} 
	\begin{array}{cc}
		1 & 0
	\end{array} \\
	\begin{array}{cc}
		0 & 1
	\end{array} \\
		\C(\psi,\phi) 
	\end{array} \right) \left(\begin{array}{c}
		v_1 \\ v_2
	\end{array}\right)
	\label{control}
\end{equation}
common to the previously referred works. In such studies $v_1$ and $v_2$ are the controlled parameters, $\psi$ and $\phi$ are typically the internal coordinates (i.e. the shape) of the locomotor, while $\Q_1$ is related to the external coordinates of the system (here, the coordinates of the centre of the cargo). The locomotion problem associated with \eqref{control} is that to determine whether a given periodic cycle of the controls $v_1$ and $v_2$ can generate a \emph{geometric phase}, that is, a net motion $\Delta \Q_1$. A geometric phase emerges if i) the controls undergo a non-reciprocal cycle and ii)  $\nabla \times \C$ is non-zero, see \cite{Coron}.

Our model lies within this general framework, with $\nabla \times \C = \V$, and one major peculiarity. Unlike all the aforementioned studies, here the control parameters cannot be chosen independently. Indeed, we have $v_1=\dot{v}_2$, since $\psi=\dot{\phi}$. However, the key requirement for obtaining a net motion is not the independence of controls, but rather their non-reciprocal evolution. Since the controls one the time derivative of the other,  non-reciprocity happens ``naturally''. The simple actuations considered before are an example of that. More generally we have
\begin{equation}
	\text{area}(\Omega)=\int_{\Omega} d \psi d \phi = \int_{\partial \Omega} \psi \, d\phi = \int_{0}^{1} \dot{\phi}^2 dt >0 \,  \label{pos}
\end{equation}
for every actuation $\phi$, where $\text{area}(\Omega)$ is the signed area of $\Omega$, namely, the area or its opposite depending on the orientation. So, for every non-trivial actuation, $\dot{\phi}$ and $\phi$ always undergo a non-reciprocal cycle. Observe also that \eqref{pos} tells that the signed area of $\Omega$ is always positive or, equivalently, a curve $t\to(\dot{\phi}(t),\phi(t))$ that parametrizes the boundary of a domain in the plane is always oriented in the counter-clockwise direction.

\subsubsection*{Direction switch}

In all the examples we have seen so far we had the swimmer moving in the negative horizontal direction, head-first. Equations \eqref{Dq} and \eqref{pos} show that this is in fact the case for \emph{every} actuation such that $|\phi|<\pi/2$. Indeed, since $\V \cdot \E_1 <0$ for every point in $\Omega$ generated by such an oscillation, and since the signed area of $\Omega$ is always positive by \eqref{pos}, the horizontal projection of the integral in \eqref{Dq} is negative. So, although the elasticity of the flagellum is crucial in order to enable locomotion, it also implies limitations on the possible direction of the swimmer. A very similar situation holds for the three-sphere swimmer with a passive elastic arm \cite{Mo}, for which locomotion is indeed possible, but only in one direction.

\begin{figure}[h]
	\centering
		\includegraphics[width=1.00\textwidth]{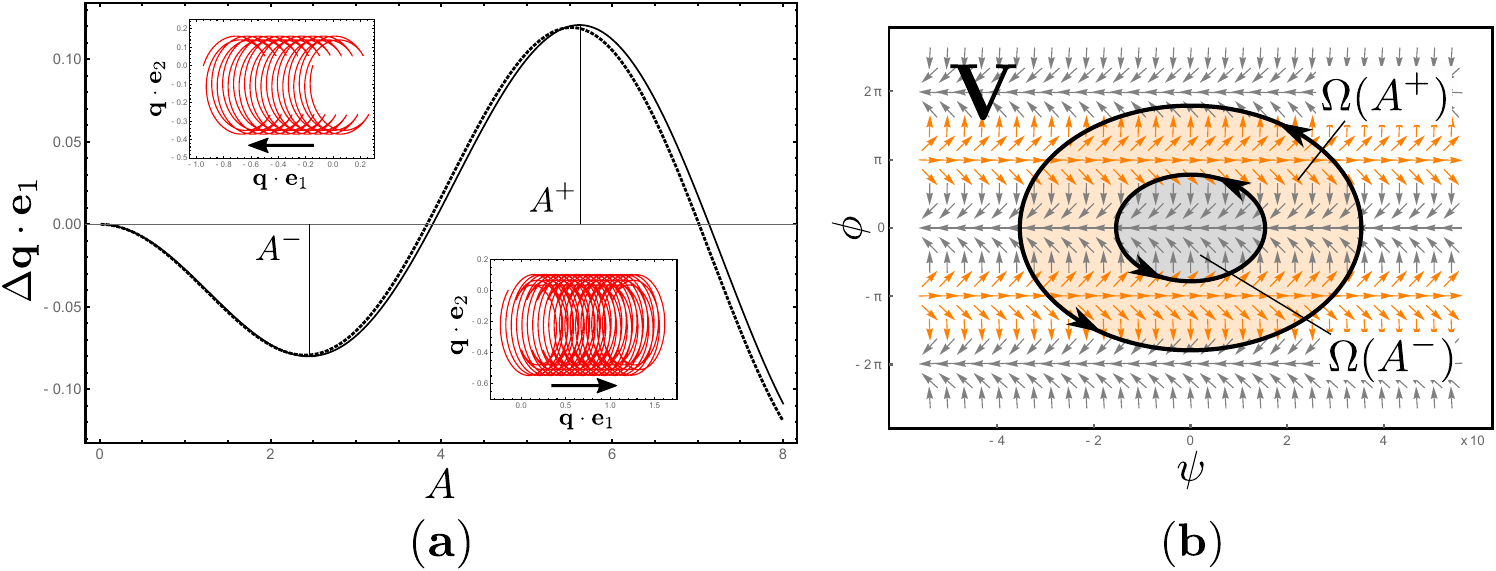}
	\caption{Sinusoidal oscillations of growing amplitude $A$ of the system with $\rho=0.15$ and $\epsilon=0.4$. (\textbf{a}) Asymptotic estimate (dotted line) and numerical solutions (solid line) for the horizontal component of the displacement as function of $A$. Two inset plots show the resulting trajectory of $\Q$ for the actuation amplitudes $A^{-}$ and $A^{+}$. (\textbf{b}) Motility map.}
	\label{fig4} 
\end{figure}

In our case, however, for actuations such that  $|\phi(t)|>\pi/2$ for some $t$, it is also possible to switch direction, because $\V \cdot \E_1 =C\cos \phi$ can change sign. Indeed, if we consider the sinusoidal actuations $\phi(t)=A \sin(2\pi t)$ and we denote by $\Omega(A)$ their generated domains, we have
\begin{equation}
	\Delta \Q \cdot \E_1 = \epsilon	\!  \int_{\Omega(A)} \V \cdot \E_1 d\psi d \phi + \Og(\epsilon^2)  = \epsilon C (2 \pi)^2 A J_{1}(A) + \Og(\epsilon^2) \label{Horiz}
\end{equation}
where $J_1$ is the first Bessel function of the first kind. We conclude that the horizontal displacement can change sign for $A$ large enough. The numerical solutions show this as well: as $A$ grows, the horizontal displacement goes from negative to positive and then negative again, passing from local maxima to local minima that grow in modulus. In Figure \ref{fig4}(\textbf{a}) we show approximated and numerical results for the displacement as function of the amplitude $A$. The motility map of the system is shown in Figure \ref{fig4}(\textbf{b}). Notice that, as $A$ grows, $\Omega(A)$ end up to gather, alternatively, more points in which $\V\cdot\E_1 <0$ or more points in which $\V\cdot\E_1>0$, thus determining the sign switch in the integral in \eqref{Horiz}.  

\subsection{Prescribed torque}\label{torque}
The natural problem for the externally actuated swimmer is the one in which we prescribe the torque $\boldsymbol\tau_{\textrm{ext}}$ acting on the cargo. The system of equations in this case is given by \eqref{ext1}-\eqref{ext3} together with the moment balance \eqref{mbal_ext}. The system can be solved numerically for unknowns $u^x$, $u^y$, $y$, and $\dot{\phi}$. We assume that the normalized external torque is of order $\Og(\epsilon)$, taking $\boldsymbol\tau_{\textrm{ext}}= \epsilon\tau_{\textrm{ext}}\E_{3}$ with $\tau_{\textrm{ext}}=\Og(1)$. Equivalently, we assume that the dimensional external torque has the same magnitude as the total viscous moment acting on the swimmer (thus, we are not restricting the amplitude of the oscillations). If $\tau_{\textrm{ext}}$ is periodic, numerical solutions are approximated by asymptotic solutions obtained with the same method adopted in the case of the prescribed angle actuation. These are given by formal series expression for $u^x$, $u^y$, and $y$ as in \eqref{asym}, together with the power series expansion for the angle
\begin{equation}
\phi=\phi_0 + \epsilon \phi_1 + \epsilon^{2} \phi_2 \ldots \label{asymphi}
\end{equation}
We provide the explicit calculations of the asymptotic coefficients in the Appendix, while we discuss here the main results.

An important formula we obtain is that for the zero order coefficient of the angle evolution $\phi(t)$, that reads 
\begin{equation}
\phi_0(t)= \phi_{\textrm{in}}+ T_{0}^{-1} \int_{0}^{t} \tau_{\textrm{ext}} \quad \text{where} \quad  T_0 = \eta_{\textrm{rot}} +  \frac{4 \eta \left(1+ 3 \rho + 3\rho^2 \right)  + 1}{12\left( \eta +1\right)} 
\label{phi_0}
\end{equation} 
and $\phi_{\textrm{in}}$ is the angle at time $t=0$. Formula \eqref{phi_0} shows that we are back to the prescribed angle case. Indeed, at leading order, the orientation of the cargo is completely determined by the prescribed torque $\tau_{\textrm{ext}}$, at least up to a multiplicative constant and an integration in the time variable.

Equation \eqref{phi_0} is combined with another result of the asymptotic calculations, namely, the expression for the displacement $\Delta \Q$. If $\tau_{\textrm{ext}}$ has zero average during one period, the formula we obtain for $\Delta \Q$ is in fact the same as in \eqref{Dq} with $\phi$ replaced by $\phi_0$, namely
\begin{equation*}
			\Delta \Q =  \epsilon \! \! \int_{\Omega} \V \, d\psi d \phi_{0}  + \Og(\epsilon^{2})
\end{equation*}
where $\V(\psi , \phi_{0})=C \E_{\phi_0}$  and $\Omega$ is the domain contained in the closed curve
\begin{equation}
\partial \Omega=\left\{(\dot{\phi}_0 (t),\phi_0 (t)) \, ; \, t \in \left[0,1\right]\right\} \subset \mathbb{R}^{2} \, . \label{partialOtorque}
\end{equation}
All the motility results deduced in Section \ref{MM_Ext} apply here with $\phi$ replaced by $\phi_0$.

\section{Internally actuated swimmer} \label{Int}
For the swimmer of Figure \ref{fig1}(\textbf{b}) we focus on the problem in which the internal angle $\alpha=\theta - \phi$ is prescribed. We restrict our analysis to the physically significant case $|\alpha| <\pi/2$.

As in the externally actuated case, we can simplify further equations \eqref{flag1}-\eqref{flag2}. Contrary to the $u^x=\Og(\epsilon)$ scaling of Section \ref{Ext}, we now have that, in general, the longitudinal velocity $u^x$ of the flagellum at the point of attachment is of order $\Og( 1)$. Consequently, we do not have $\sigma =\Og(\epsilon^2)$. However, using \eqref{BC} and \eqref{flag1} it is easy to conclude that
\begin{displaymath}
	\sigma = \epsilon \g u^{x}(x-1) + \Og(\epsilon^2), 
\end{displaymath}
assuming  again that $y$ and its derivatives are of order $\Og(\epsilon)$. Just as in the externally actuated case, we drop terms of order $\Og(\epsilon^3)$ in equation \eqref{flag2}. We now obtain
\begin{equation}
\epsilon \left( u^y + x\dot{\theta} + \frac{\partial y}{\partial t} +(\g - 1)u^{x}\frac{\partial y}{\partial x}\right)  =  - \frac{\partial^4 y}{\partial x^4} + \epsilon \frac{\partial}{\partial x}\left(\g u^{x}(x-1) \frac{\partial y}{\partial x}\right) \, ,
\label{int1}
\end{equation}
where $\theta=\phi+\alpha$. Observe that  $y$ and $\sigma$ are again decoupled. Applying definitions \eqref{uxuy} and integrating \eqref{flag1} using \eqref{BC}, we can rewrite equations \eqref{fbal} and \eqref{mbal_int} as
\begin{empheq}[left=\empheqlbrace]{align}
& \: - \epsilon(\eta +\g )u^x + \epsilon \eta \rho \dot{\phi} \sin \alpha + \epsilon \g  \int_{0}^{1} \! \dot{\theta} y \,dx  - \epsilon (\g-1) \! \int_{0}^{1} \! (u^y + x \dot{\theta})\frac{\partial y}{\partial x} \, dx = 0 \label{int2} \\
& \: - \epsilon \eta u^{y}(t) + \epsilon \eta \rho \dot{\phi} \cos \alpha   - \frac{\partial^3 y}{\partial x^3}(0,t) = 0 \label{int3} \\
& \: - \epsilon (\eta_{\text{rot}} + \eta \rho^2 )\dot{\phi} + \epsilon \eta \rho u^x \sin \alpha + \epsilon \eta \rho u^y \cos \alpha  + \frac{\partial^2 y}{\partial x^2}(0,t) = 0 \label{int4}
\end{empheq}
System \eqref{int1}-\eqref{int4}, together with boundary conditions \eqref{attach} and \eqref{BC}, defines a set of four equations in the unknowns $u^x$, $u^y$, $\dot{\phi}$, and $y$. We solve numerically these equations using again a finite difference scheme based on that in \cite{She}. Given a periodic $\alpha$, solutions approach periodic orbits after a brief transient. As for the externally driven case, the periodic long time behaviour can be approximated using standard series expansion methods. 

\subsection{Asymptotics} \label{Asym_Int}
We look again for formal solutions where $u^x$, $u^y$, and $y$ are given by \eqref{asym}. While writing also $\phi$ and $\theta$ as power series of $\epsilon$, we assume that their coefficients satisfy $\theta_{0}= \phi_{0} + \alpha$ and $\theta_{k}=\phi_{k}$ for $k\geq 1$. We provide here only the main steps to compute the expansions up to order $k=1$.

As in the externally actuated case we have $y_{0}=0$, while the equation for $y_1$ reads as \eqref{eq_y1}  with $\dot{\phi}_{0} +\dot{\alpha}$ instead of $\dot{\phi}$. Boundary conditions for $y_1$ are given by \eqref{BCk}, thus we can write
\begin{equation}
y_{1}(x,t) = - \int_{0}^{x} \! \int_{0}^{x_1} \! \! \int_{x_2}^{1} \! \int_{x_3}^{1} \Big( u^{y}_{0}(t) + (\dot{\phi}_{0}(t) + \dot{\alpha}(t))x_4 \Big) \,  dx_{4}  dx_{3}  dx_{2}  dx_{1}  \, .\label{y1_int}  	
\end{equation}
We have then
\begin{equation*}
\frac{\partial^3 y_1}{\partial x^3}(0,t) = u_{0}^y(t) + \frac{1}{2}(\dot{\phi}_0(t) + \dot{\alpha}(t)) \quad  \textrm{and} \quad  \frac{\partial^2 y_1}{\partial x^2}(0,t)=- \frac{u_{0}^y(t)}{2} -\frac{1}{3}(\dot{\phi}_0(t) + \dot{\alpha}(t)).
\end{equation*}
Notice that the first order expansion of the left hand side of \eqref{int2} is given by $-(\eta + \g)u_{0}^x  +  \eta \rho \dot{\phi}_{0} \sin \alpha$. Therefore, the first order in $\epsilon$ of \eqref{int2}-\eqref{int4} gives
\begin{displaymath}
	M(\alpha)\left( \begin{array}{c} u_{0}^{x} \\ u_{0}^{y} \\  \dot{\phi}_{0} \end{array}\right) = \left( \begin{array}{c} 0 \\ \dot{\alpha}/2 \\  \dot{\alpha}/3 \end{array}\right)  
\end{displaymath}	
where 
\begin{equation}	
	M(\alpha) = \left(\begin{array}{ccc} -(\eta + \g) & 0 & \eta \rho \sin \alpha \\
	0 & -(\eta +1) &\eta \rho \cos \alpha - 1/2 \\
	\eta \rho \sin \alpha  & \eta \rho \cos \alpha -1/2 &-( \eta_{\text{rot}} + \eta \rho^2 + 1/3 ) 		 
	\end{array}\right) \, . \label{M}
\end{equation}
By inverting $M$ we obtain that $u_{0}^{x}$, $u_{0}^{y}$, $\dot{\phi}_{0}$ and $\dot{\theta}_{0}$ can be written as
\begin{equation}
u_{0}^{x}=U_{0}^{x}(\alpha) \dot{\alpha}	\, , \: \:  u_{0}^{y}=U_{0}^{y}(\alpha) \dot{\alpha}  \, , \: \: \dot{\phi}_{0} = \p_{0}(\alpha) \dot{\alpha} \, , \: \: \: \textrm{and} \: \:  \: \: \dot{\theta}_{0}= \vartheta_{0}(\alpha) \dot{\alpha}  
\label{uxuyalpha}
\end{equation}
where $U_{0}^{x}$, $U_{0}^{y}$ and $\p_{0}$ are functions of $\alpha$ that can be calculated explicitly, while $\vartheta_{0}=\p_{0}  + 1$. Taking two primitives $\Phi_0$ and $\Theta_0$, respectively for $\p_0$ and $\vartheta_0$, that are compatible with the initial conditions, we obtain
\begin{equation}
\phi_{0}=\Phi_0(\alpha) \quad \textrm{and} \quad \theta_{0}=\Theta_0(\alpha) \, .
\label{phitheta}
\end{equation}
In turn, we have that 
\begin{equation}
\begin{split}
& \quad \quad  \quad \quad  y_1(x,t)=-p_1(x,\alpha)\dot{\alpha} \quad \textrm{where}  \\
& p_1(x,\alpha)=\int_{0}^{x} \! \int_{0}^{x_1} \! \! \int_{x_2}^{1} \! \int_{x_3}^{1} \Big( U^{y}_{0}(\alpha) + \frac{d \Theta_{0}}{d \alpha}(\alpha) x_4 \Big) \,  dx_{4}  dx_{3}  dx_{2}  dx_{1}
\end{split}
\label{p1alpha}
\end{equation}
is a polynomial in $x$ with $\alpha$-dependent coefficients. Notice that the bending of the flagellum, at leading order, depends only on the internal angle $\alpha$ and its velocity $\dot{\alpha}$. 

Now, the second order expansion of \eqref{int1} reads
\begin{equation}
u_{1}^y + x\dot{\theta}_{1} + \frac{\partial y_1}{\partial t} +(\g - 1)u_{0}^{x}\frac{\partial y_1}{\partial x}  =  - \frac{\partial^4 y_2}{\partial x^4} +  \frac{\partial}{\partial x}\left(\g u_{0}^{x}(x-1) \frac{\partial y_1}{\partial x}\right) \, . \label{y2_Int}
\end{equation}
The only solution for $y_2$ solving \eqref{y2_Int} with boundary conditions \eqref{BCk} is
\begin{equation*}
\begin{split}
y_{2}(x,t)  = & - \int_{0}^{x} \! \int_{0}^{x_1} \! \! \int_{x_2}^{1} \! \int_{x_3}^{1} \Big( u^{y}_{1}(t) + \dot{\phi}_{1}(t) x_4 + \frac{\partial y_1}{\partial t}(x_4,t)  \Big) \,  dx_{4}  dx_{3}  dx_{2}  dx_{1}  \\
            + &  \int_{0}^{x} \! \int_{0}^{x_1} \! \! \int_{x_2}^{1} \! \int_{x_3}^{1} u^{x}_{0}\Big( \frac{\partial y_1}{\partial x}(x_4,t) + \gamma (x_{4}-1)\frac{\partial^2 y_1}{\partial x^2}(x_4,t)  \Big) \,  dx_{4}  dx_{3}  dx_{2}  dx_{1} \, .
\end{split}
\end{equation*}
From the previous results we have that $y_2$ is a polynomial in $x$ with coefficients depending on $u^{y}_{1}$ and $\dot{\phi}_{1}$, together with functions of $\alpha$ multiplied by either $\dot{\alpha}^2$ or $\ddot{\alpha}$. In particular we have
\begin{equation*}
\begin{split}
& \frac{\partial^3 y_2}{\partial x^3}(0,t) = u_{1}^y(t) + \frac{\dot{\phi}_1(t)}{2} - a_2(\alpha)\dot{\alpha}^2 - b_2(\alpha)\ddot{\alpha} \\
& \frac{\partial^2 y_2}{\partial x^2}(0,t) = - \frac{u_{1}^y(t)}{2} -\frac{\dot{\phi}_1(t)}{3} + a_3(\alpha)\dot{\alpha}^2  + b_3(\alpha)\ddot{\alpha}  \, ,
\end{split}
\end{equation*}
where the $\alpha$-dependent functions $a_2$, $b_2$, $a_3$, and $b_3$ can be calculated explicitly. In the same fashion we can write the second order term in the expansion of \eqref{int2} as
\begin{equation*}
\begin{split}
 & - (\eta +\g )u_{1}^x + \eta \rho \dot{\phi}_{1} \sin \alpha +  \g  \int_{0}^{1} \! \dot{\theta}_{0} y_1 \,dx  -  (\g-1) \! \int_{0}^{1} \! (u_{0}^y + x \dot{\theta}_{0})\frac{\partial y_{1}}{\partial x} \, dx \\
  & \quad \quad = - (\eta +\g )u_{1}^x + \eta \rho \dot{\phi}_{1} \sin \alpha -a_{1}(\alpha)\dot{\alpha}^2  - b_{1}(\alpha)\ddot{\alpha} \, .
\end{split}
\end{equation*}
Combining these results, the second order expansion of \eqref{int2}-\eqref{int4} gives
\begin{displaymath}
	M(\alpha)\left( \begin{array}{c} u_{1}^{x} \\ u_{1}^{y} \\  \dot{\phi}_{1} \end{array}\right) =
	\left(\begin{array}{c}  a_{1}(\alpha) \\
 a_2(\alpha)\\ 
a_3(\alpha)\end{array}\right) \dot{\alpha}^2 + 
	\left(\begin{array}{c} b_{1}(\alpha) \\
 b_{2}(\alpha) \\ 
b_3(\alpha)\end{array}\right) \ddot{\alpha}
\end{displaymath}
where $M$ is given again by \eqref{M}. By inverting $M$ we obtain  $u_{1}^{x}$, $u_{1}^{y}$, and $\dot{\phi}_{1}=\dot{\theta}_1$ which can be written in the following form
\begin{eqnarray}
& u_{1}^{x} = A^{x} (\alpha) \dot{\alpha}^2 + B^{x} (\alpha) \ddot{\alpha} & \label{ABux}\\
& u_{1}^{y} = A^{y} (\alpha) \dot{\alpha}^2 + B^{y} (\alpha) \ddot{\alpha} & \label{ABuy}\\
& \dot{\phi}_{1}  = A^{\phi} (\alpha) \dot{\alpha}^2 + B^{\phi} (\alpha) \ddot{\alpha} \, ,\label{ABphi}&
\end{eqnarray} 
where all the functions of $\alpha$ in the right hand sides can be calculated explicitly.

\subsection{Swimming trajectories and motility maps}\label{MM_Int}
We discuss here the behaviour of the orientation $\phi$ and  the coordinate $\Q$ of the cargo, by applying the asymptotic results of the previous section. For the sake of simplicity we assume that solutions of \eqref{int1}-\eqref{int4} are periodic for $t\geq 0$. By integrating $\dot{\phi}$ and 
\begin{equation}
\dot{\Q}= \frac{d}{dt} \Big( \R(0,t) - \rho \E_{\phi} \Big)=  u^{x} \E_{\theta} + u^{y}\E_{\theta}^{\bot} - \rho \dot{\phi} \E_{\phi}^{\bot}
\label{qdot_int}
\end{equation}
we show that the swimmer can i) vary its orientation $\phi$, and ii) move along curved trajectories.

\begin{figure}[h]
	\centering
		\includegraphics[width=1.0\textwidth]{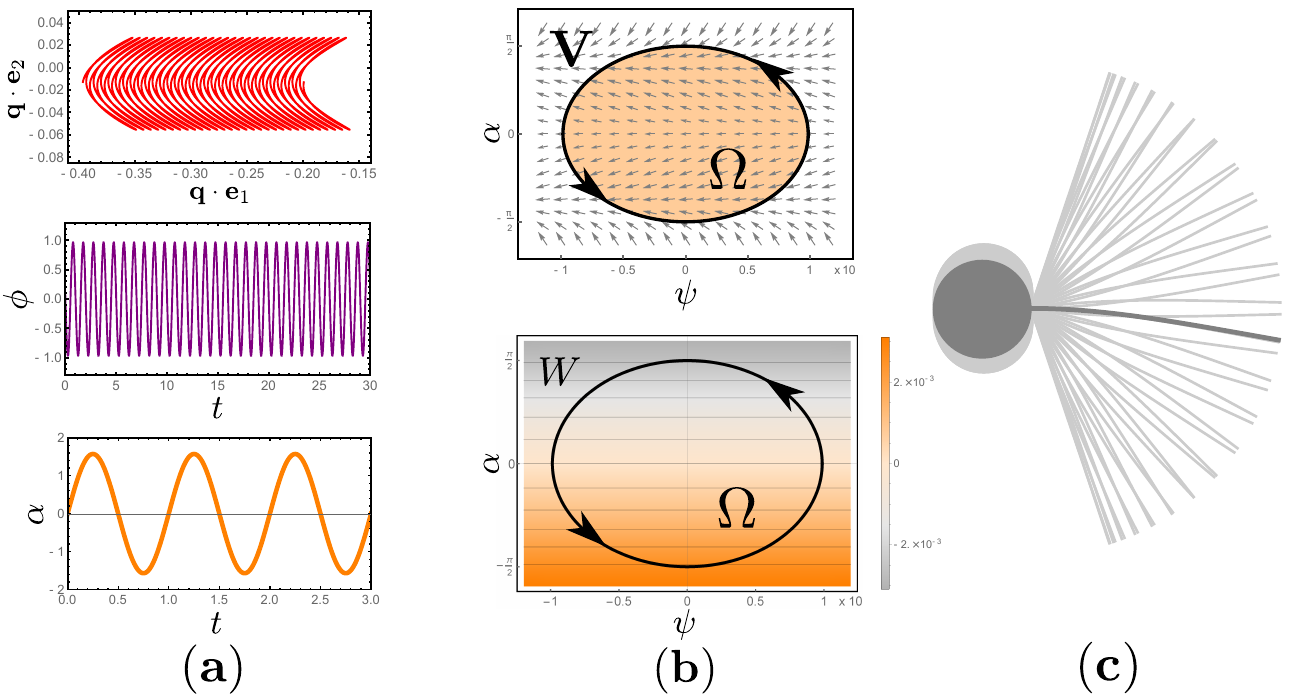}
	\caption{Sinusoidal oscillations of the internally actuated swimmer with $\rho=0.2$ and $\epsilon=0.7$. (\textbf{a}) The actuation $\alpha$ (bottom) and the numerical solution for $\phi$ (middle) and $\Q$ (top).  (\textbf{b}) Motility maps. (\textbf{c}) Snap-shots of the swimmer in motion, as seen by a frame moving with the contact point between cargo and flagellum.}
	\label{fig5}
\end{figure}

We start by considering the orientation $\phi$ of the cargo. Since $\dot{\phi}$ is periodic, the net variation of the angle $\phi$ over \emph{any} period of the actuation $n \geq 1$ is given by
\begin{displaymath}
	\Delta \phi := \int_{n}^{n+1} \dot{\phi} = \int_{0}^{1} \dot{\phi} \, .
\end{displaymath}
Expanding the solution we have $\Delta \phi=\Delta \phi_0 +\epsilon\Delta \phi_1 + \Og(\epsilon^2)$. Using \eqref{phitheta}, we obtain $\Delta \phi_0 = \left[\Phi_0(\alpha)\right]_{0}^{1}=0$. Then, we notice that formula \eqref{ABphi} for $\dot{\phi}_1$ has the form \eqref{AB}, with $\phi$ replaced by $\alpha$. Following the same arguments of Section \ref{MM_Ext}, we obtain 
\begin{equation}
\Delta \phi_1  =\int_{\Omega} W  \, d\psi d\alpha  \quad \textrm{where} \quad W(\psi , \alpha )=A^{\phi}(\alpha)-\frac{dB^{\phi}}{d\alpha}(\alpha)
\label{Dphi1}
\end{equation}
can be explicitly calculated, and $\Omega$ is a domain in $\mathbb{R}^{2}$ such that 
\begin{equation}
\partial \Omega=\left\{(\dot{\alpha}(t),\alpha(t)) \, ; \, t \in \left[0,1\right]\right\} \subset \mathbb{R}^{2} \, . \label{boundaryalpha}
\end{equation}
Calculations show that $W$ is non-zero for every $\rho>0$, see Figure \ref{fig5}. As a result, the net variation of the angle 
\begin{equation}
	\Delta \phi  = \epsilon \int_{\Omega} W \, d\psi d\alpha + \Og(\epsilon^2)
\label{Dphi}
\end{equation}
can be non-zero, and the swimmer can change its orientation. 

Let us consider now the coordinate $\Q$ of the cargo. Observe first that $\phi(t+n)=\phi(t) +  n \Delta \phi$. Then, the displacement $\Delta_n \Q$ at the $n^{\textrm{th}}$ actuation cycle is
\begin{equation}
\begin{split}
\Delta_n \Q \:  & = \int_{n}^{n+1}\dot{\Q} \: \: =  \: \int_{n}^{n+1}\left( u^{x} \E_{\theta} + u^{y}\E_{\theta}^{\bot} - \rho \dot{\phi}\E_{\phi}^{\bot}\right)   \\
&= \int_{0}^{1}\Big( u^{x} \E_{\theta + n\Delta \phi} + u^{y}\E_{\theta +n\Delta \phi}^{\bot}  - \rho \dot{\phi}\E_{\phi+n\Delta \phi}^{\bot} \Big)  \\
&= \: \: \Rot(n\Delta \phi) \int_{0}^{1}\Big( u^{x} \E_{\theta} + u^{y}\E_{\theta}^{\bot}  - \rho \dot{\phi}\E_{\phi}^{\bot} \Big) \: \: = \: \: \Rot(\Delta \phi)^{n} \Delta_0 \Q 
\label{delta_n}
\end{split}
\end{equation}
where $\Rot(\Delta \phi)$ is the rotation matrix by the angle $\Delta \phi$. If $\Delta \phi \neq 0$ and $\Delta_{0} \Q \neq \mathbf{0}$ the swimmer moves along a curve passing through the points $\sum_{i}^{n}\Delta_i \Q $ with $n\geq 0$. In the following we deduce the first order approximation for $\Delta_{0} \Q= \Delta_{0} \Q_0 + \epsilon \Delta_{0} \Q_1 +\Og(\epsilon^2)$. For the sake of simplicity we consider here actuations $\alpha$ such that $\alpha(0)=0$. The following construction can be modified accordingly for the general case.

Expanding $\dot{\Q} = \dot{\Q}_{0} + \epsilon \dot{\Q}_1 + \Og(\epsilon^2)$, from  \eqref{uxuyalpha} and \eqref{phitheta} we have that
\begin{equation*}
	\dot{\Q}_{0}  = (u_{0}^{x} \E_{\theta_0} + u_{0}^{y} \E_{\theta_0}^{\bot}) - \rho \dot{\phi}_{0}\E_{\phi_0}^{\bot}
\end{equation*}
has an expression of the form $\mathbf{P}(\alpha) \dot{\alpha}$, which gives always a null result when integrated over a period. Therefore $\Delta_{0}\Q_{0}=0$. The first order coefficient in the expansion for $\dot{\Q}$ reads
\begin{equation*}
	\dot{\Q}_{1}  = \underbrace{(u_{1}^{x} \E_{\theta_0} + u_{1}^{y} \E_{\theta_0}^{\bot}) }_{I}
	+  \underbrace{\phi_1 (u_{0}^{x} \E_{\theta_0}^{\bot} - u_{0}^{y} \E_{\theta_0}) }_{II}
	- \underbrace{\rho\dot{\phi}_1 \E_{\phi_0}^{\bot}}_{III} 
	+ \underbrace{\phi_1 \rho  \dot{\phi}_0\E_{\phi_0}}_{IV} \, .
\end{equation*}
Now, from \eqref{phitheta} and \eqref{ABux}-\eqref{ABphi} we see that terms I and III have the form \eqref{AB}. Using Stokes theorem they can be written, respectively, as integrals of two vector fields $\V_{I}$ and $\V_{III}$ over $\Omega$, where the boundary $\partial \Omega$ is given by \eqref{boundaryalpha}. On the other hand, terms II and IV are products of $\phi_1$ times an expression of the type $\mathbf{P}(\alpha) \dot{\alpha}$. If we choose a function $\mathbf{\Pi}_{0}(\alpha)$ with $d\mathbf{\Pi}_{0}/d\alpha= \mathbf{P}$ such that $\mathbf{\Pi}_{0}(0)=\mathbf{0}$, then we have
\begin{displaymath}
	\int_{0}^{1} \phi_{1} \mathbf{\Pi}_{0}'(\alpha) \dot{\alpha} = \left[\phi_{1} \mathbf{\Pi}_{0}(\alpha)\right]_{0}^{1} - \int_{0}^{1} \dot{\phi}_{1} \mathbf{\Pi}_{0}(\alpha) = -  \int_{0}^{1} \dot{\phi}_{1} \mathbf{\Pi}_{0}(\alpha) \, .
\end{displaymath}
Since  $\dot{\phi}_{1} \mathbf{\Pi}_{0}(\alpha)$ has now the form \eqref{AB}, then  II and IV can be written, respectively, as integrals of two vector fields $\V_{II}$ and $\V_{IV}$ over $\Omega$. Taking $\V=\V_{I}+\V_{II}+\V_{III}+\V_{IV}$ we have that
\begin{equation}
\Delta_{0} \Q_1 = \int_{\Omega} \V \, d \psi d \alpha \, .
\label{Dq1alpha}
\end{equation}

Summarizing, the net rigid motion of the swimmer after the $n^{\text{th}}$ cycle is given by the rotation $\Rot(\Delta \phi)$ and the translation \eqref{delta_n}. The asymptotic solutions give us an approximation for the net rotation and translation through formulas \eqref{Dphi} and 
\begin{equation}
\Delta_{0} \Q = \epsilon \int_{\Omega} \V \, d \psi d \alpha \,  + \Og(\epsilon^2) \, ,
\label{Dq_int}
\end{equation}
which allow for a motility analysis solely based on the geometry of $\Omega$, since $W$ and $\V$ do not depend on the given actuation $t \mapsto \alpha(t)$.

There are some similarities and some  differences between the results we have deduced here and the ones of Section \ref{Ext}. As for the externally actuated swimmer, the ``control'' parameters are given by the actuation, $\alpha$ in this case, and its velocity. Observe that from formula \eqref{p1alpha} we have $y(\cdot)= -\epsilon p_1 (\alpha,\cdot)\dot{\alpha} + \Og(\epsilon^2)$. The bending $y$ is proportional to $\dot{\alpha}$ again because the forces acting on the flagellum are of viscous nature. Thus, at leading order, $\alpha$ and $\dot{\alpha}$ determine completely the shape of the whole swimmer (cargo+flagellum). Calculations shows that $\V$ is non zero. Then, from formula \eqref{Dq_int}, we recover the celebrated ``Scallop Theorem'': in order to produce net advancement, the shape of the swimmer must undergo non-reciprocal cycles. From \eqref{pos}, we have that this is always the case for any non-trivial actuation. 

Notice that, despite being driven by a single input, the internally actuated swimmer is able to change not only is position, but also its orientation $\phi$. However, translations and orientation are correlated: we can not control $\Delta_{n} \Q$ and $\Delta\phi$ independently.

\subsubsection*{Straight and curved trajectories}

Calculations show that $W(\psi,\alpha)=W(\alpha)$ is odd with respect to $\alpha$ and negative for $\alpha>0$. From \eqref{Dphi} we have that for domains $\Omega$ that are symmetric with respect to the $\psi$ axis there are, at leading order, no net rotations of the swimmer after any cycle. This happens, for example, with a sinusoidal actuation $\alpha(t)= \pi/2 \sin(2\pi t)$. Since $\V \cdot \E_1 <0$ for every $|\alpha|<\pi/2$ while $\V\cdot \E_2$ is odd with respect to  $\alpha$, the swimmer moves head-first and, on average, on the horizontal axis. We recover here the same swimming behaviour described by previous studies \cite{Lauga,Or1}.

A second result we deduce from \eqref{Dphi} is that the swimmer rotates when it performs sinusoidal beating of the flagellum $\alpha(t)=\alpha^* + A \sin(2\pi t)$ around a non-zero internal angle $\alpha^*$. The sign of $W$ leads to a counter-clockwise rotation of the swimmer for $\alpha^*<0$ and a clockwise rotation for $\alpha^*>0$. While  (to the best of our knowledge) this effect has never been discussed before, it does not rely on large actuation amplitudes, and it is also present in the small-actuation regime considered in \cite{Lauga}.

\begin{figure}[h]
	\centering
		\includegraphics[width=1.00\textwidth]{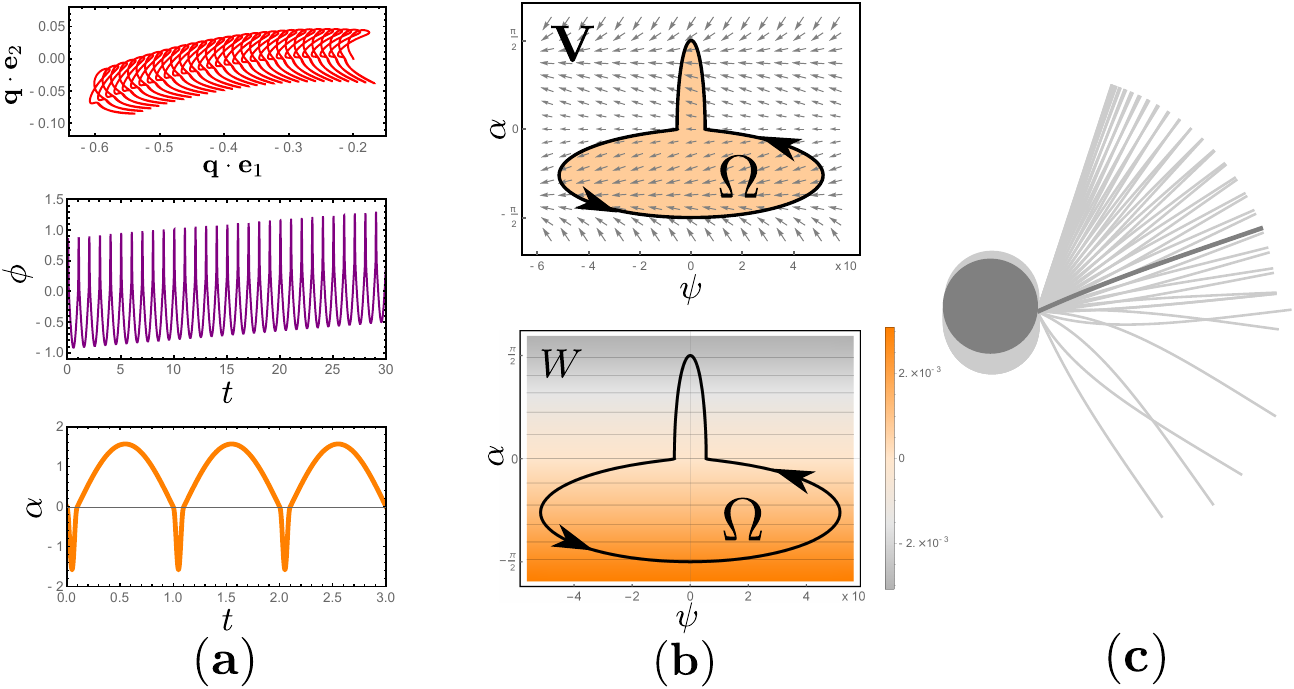}
		\caption{Non-sinusoidal oscillations of the internally actuated swimmer with $\rho=0.2$ and $\epsilon=0.7$. (\textbf{a}) The actuation $\alpha$ (bottom) and the numerical solution for $\phi$ (middle) and $\Q$ (top). (\textbf{b}) Motility maps. (\textbf{c}) Snap-shots of the swimmer in motion, as seen by a frame moving with the contact point between cargo and flagellum.}
	\label{fig6}
\end{figure}

The novel physical insight we get from \eqref{Dphi} and \eqref{Dq_int} is related to non-sinusoidal actuations, when $\Omega$ is non-symmetric with respect to the $\psi$ axis. We consider here the actuation in Figure \ref{fig6}(\textbf{a}), for which $\dot{\alpha}$ is much larger when $\alpha<0$. The part of the resulting domain $\Omega$ where $W$ is positive is then  larger than the part of $\Omega$ where $W$ is negative. As a result the swimmer rotates counter-clockwise at every actuation cycle. The displacement $\Delta_{0}\Q$ is, at leading order, directed in the negative horizontal direction, since $\V \cdot \E_1 <0$ for every point in the space $(\psi,\alpha)$. The composition of rotations and translations at every cycle generates the curved path illustrated in Figure \ref{fig6}(\textbf{a}).

\section{Conclusions and future work}
We have studied the locomotion capabilities of the flagellar microwimmers of Figure \ref{fig1}(\textbf{a}) and \ref{fig1}(\textbf{b}), for generic periodic actuations. Our approach relies on the assumption \eqref{epsilon_fla} of small compliance  of the flagella, which allow us to perform an explicit asymptotic analysis leading to explicit formulas and motility maps. These are further validated by numerical simulations. We have shown that, by modulating the velocity of the inputs, the externally actuated swimmer can translate laterally with respect to this symmetry axis of beating, while the internally actuated one is able to move along curved trajectories. Moreover, we found a direction switch in the average velocity for large enough amplitude of oscillations for the externally actuated swimmer.

A possible direction for future research on these models is to consider flagella of arbitrary stiffness, thus dropping the hypothesis \eqref{epsilon_fla}. For floppy flagella,  travelling bending waves are expected, together with increased swimming speed and efficiency.  To the best of our knowledge, the question whether lateral displacements can be induced by modulating the actuation velocity has not yet been addressed. Moreover, it would be interesting to explore whether  velocity modulation can provide a steering mechanism also in different (yet related) systems, like swimmers with flagella actuated by distributed internal torque. All of these questions will require further study.

\vspace{1cm}
\noindent \textbf{Acknowledgements} We gratefully acknowledge the support by the European Research Council through the ERC Advanced Grant 340685-MicroMotility.

\section*{Appendix. Prescribed external torque: \\ formal asymptotic solution.}

The governing equations and assumptions for the swimmer driven by an external torque are listed in the beginning of Section \ref{torque}. As for the prescribed angle case, $u^x$ is decoupled from the other unknowns of the problem. We take $u^x$, $u^y$, $y$ as in \eqref{asym} and $\phi$ as in \eqref{asymphi}.  The coefficients of $u^y$, $y$,  and $\phi$ must solve \eqref{ext1k}-\eqref{ext2k} and the boundary conditions \eqref{BCk}. Moreover, expanding equation \eqref{mbal_ext} we have
\begin{equation}
 \frac{\partial^2 y_0}{\partial x^2}(0,t)=0 \quad \text{and} \quad - \eta_{\textrm{rot}}  \dot{\phi}_k + \eta\rho(u_{k}^{y}-\rho\dot{\phi}_k) + \frac{\partial^2 y_{k+1}}{\partial x^2}(0,t) + \tau_k =0   \label{ext4k}
\end{equation}
for $k\geq 0$, where we take $\tau_0 =\tau_{\textrm{ext}}$ and $\tau_k =0$ for $k \geq 1$. Then, each coefficient of $u^x$ can be directly calculated expanding  \eqref{ext3}. If the initial value $\phi_{\textrm{in}}$ for the angle at $t=0$ is prescribed, we impose $\phi_0(0)= \phi_{\textrm{in}}$ and  $\phi_k(0)=0$ for $k\geq 0$.

At  order zero we have $y_{0}(x,t)=0$, therefore $y_1$ can be written as \eqref{y1} with $\dot{\phi}_0$ instead of $\dot{\phi}$. Substituting this expression for $y_1$ in \eqref{ext2k} and \eqref{ext4k} we obtain the following linear system
\begin{empheq}[left=\empheqlbrace]{align}
	& \: \eta \rho \dot{\phi}_0 -  \eta u_{0}^{y} =  u^{y}_{0} + \frac{1}{2}\dot{\phi}_0 \nonumber \\
	& \: - \eta_{\textrm{rot}}  \dot{\phi}_0 + \eta\rho(u_{0}^{y}-\rho\dot{\phi}_0) - \frac{1}{3}\dot{\phi}_0 -\frac{1}{2}u^{y}_{0}  + \tau_{\textrm{ext}} =0 \nonumber
\end{empheq}
which allows us to solve for $u^{y}_{0}$ and $\dot{\phi}_0$ in terms of $\tau_{\textrm{ext}}$. If we consider the constants $U_{0}^{y}=(\eta \rho +1/2)/(\eta+1)$ and $T_{0}$ as in  \eqref{phi_0}, then we have
\begin{displaymath}
	 u_{0}^{y}(t) = U_{0}^y \, \dot{\phi}_0 (t) \quad \textrm{and} \quad \dot{\phi}_0 (t) = T_{0}^{-1} \tau_{\textrm{ext}}(t)
\end{displaymath}
which gives the formula for $\phi_0$ in \eqref{phi_0}. In turn $y_{1}(x,t) =  -p_1(x) \dot{\phi}_0 (t)$ with $p_1$ given by \eqref{p1}, thus the $k=2$ order problem in \eqref{ext1k} is solved by
\begin{displaymath}
	y_{2}(x,t) = - \int_{0}^{x} \! \int_{0}^{x_1} \! \! \int_{x_2}^{1} \! \int_{x_3}^{1} \Big( u^{y}_{1}(t) + \dot{\phi}_1 (t) x_4 - p_1(x_4)\ddot{\phi}_0 (t) \Big) \,  dx_{4}  dx_{3}  dx_{2}  dx_{1} \, .
\end{displaymath}
Substituting the previous formula in \eqref{ext2k} and \eqref{ext4k} we obtain
\begin{empheq}[left=\empheqlbrace]{align}
	& \: \eta \rho \dot{\phi}_1 -  \eta u_{1}^{y} =  u^{y}_{1} + \frac{1}{2}\dot{\phi}_1 - \left(\int_{0}^{1}p_1\right) \ddot{\phi}_0  \nonumber\\
	& \: - \eta_{\textrm{rot}}  \dot{\phi}_1 + \eta\rho(u_{1}^{y}-\rho\dot{\phi}_1) - \frac{1}{3}\dot{\phi}_1 -\frac{1}{2}u^{y}_{1}  + \left(\int_{0}^{1} \! \! \int_{x}^{1}p_1\right) \ddot{\phi}_0 = 0 \nonumber
\end{empheq}
By defining the constants
\begin{equation*}
\p_1= T_{0}^{-1}\left( \int_{0}^{1} \! \! \int_{x}^{1}p_1 +  \frac{\eta \rho - \frac{1}{2}}{\eta +1} \int_{0}^{1}p_1 \right) \quad \textrm{and} \quad \widetilde{U}_{1}^{y} = U_{0}^{y} \p_1 + U_{1}^{y} \, ,
\end{equation*}
where $U_{1}^y$ is taken as in \eqref{uy1}, the solution of the previous system reads
\begin{displaymath}
	u_{1}^{y}(t) = \widetilde{U}_{1}^{y} \, \ddot{\phi}_0 \quad \textrm{and} \quad \dot{\phi}_1 (t) = \p_1 \ddot{\phi}_0 (t) \, .
\end{displaymath}
If we now expand the right hand side of \eqref{ext3} we obtain
\begin{displaymath}
	\epsilon(\eta+\g) u^{x} =  \epsilon^2 \g \int_{0}^{1} \dot{\phi}_0  y_1 +  \epsilon^2 \frac{\g -1}{2} \left(\frac{\partial^2 y_1}{\partial x^2}\right)^2 \! \! (0,t) + \Og(\epsilon^3) \, ,
\end{displaymath}
therefore we have $u_{0}^{x}=0$ and $u_{1}^{x} =U_{1}^{x} \dot{\phi}_{0}^{2}$, where $U_{1}^x$ is given as in \eqref{ux}. In the following we assume that $\tau_{\textrm{ext}}$ has zero average, thus $\phi_0$ is periodic. For simplicity we also assume $\tau_{\textrm{ext}}(0)=0$ so that $\phi_1(t)=\p_1 \dot{\phi}_{0}(t)$, but the result we propose here can be generalized for any initial value  of the external torque.

Let us consider the expansion $\dot{\Q}=\dot{\Q}_0 + \epsilon \dot{\Q}_1 + \Og(\epsilon^2)$. At order zero we have
\begin{equation*}
\dot{\Q}_0 =  u_{0}^x \, \E_{\phi_{0}} + u_{0}^y \, \E_{\phi_{0}}^{\bot} - \rho \dot{\phi}_{0} \, \E_{\phi_0}^{\bot} = \left(U_{0}^{y} - \rho\right)\dot{\phi}_{0} \, \E_{\phi_0}^{\bot}
\end{equation*}
while at the first order
\begin{equation*}
\begin{split}
\dot{\Q}_1 & =   \left( u_{1}^x \, \E_{\phi_{0}} + u_{1}^y \, \E_{\phi_{0}}^{\bot} + u_{0}^x \phi_{1} \, \E_{\phi_{0}}^{\bot} - u_{0}^y \phi_{1} \, \E_{\phi_{0}} - \rho \dot{\phi}_{1} \, \E_{\phi_0}^{\bot} + \rho \dot{\phi}_{0} \phi_{1} \, \E_{\phi_0} \right)    \\
& = \left(U_{1}^{x} -U_{0}^{y}\p_1 + \rho \p_1 \right) \dot{\phi}^{2}_{0} \E_{\phi_{0}} +\left(\widetilde{U}_{1}^{y} - \rho \p_1 \right)\ddot{\phi}_{0} \E_{\phi_{0}}^{\bot} \, .
\end{split}
\end{equation*}
If we now expand $\Delta \Q=\Delta \Q_{0} + \epsilon \Delta \Q_1 + \Og(\epsilon^2)$, then we have
\begin{displaymath}
			\Delta \Q_0 =  \int_{0}^{1}\left(U_{0}^{y} - \rho\right)\dot{\phi}_{0} \, \E_{\phi_0}^{\bot}=\left(U_{0}^{y} - \rho\right) \left[\E_{\phi_0}\right]_{0}^{1} = 0
\end{displaymath}
because of the periodicity of $\phi_0$. On the other hand, since $\dot{\Q}_1$ has the form \eqref{AB} with $\A(\phi_0)=\left(U_{1}^{x} -U_{0}^{y}\p_1 + \rho \p_1 \right) \E_{\phi_{0}}$ and $\B(\phi_0)=\left(\widetilde{U}_{1}^{y} - \rho \p_1 \right)\E_{\phi_{0}}^{\bot}$, if we apply \eqref{stokes} we obtain
\begin{displaymath}
	\Delta \Q_1 = \int_{\Omega} \left(\A(\phi_0) - \frac{d \B}{ d \phi_{0}}(\phi_{0})\right)  d \psi d\phi_0= (U_{1}^{x} -U_{0}^{y}\p_1 + \widetilde{U}_{1}^{y}) \int_{\Omega} \E_{\phi_0}  d \psi d\phi_0
\end{displaymath}
where $\partial \Omega$ is given by \eqref{partialOtorque}. Since $U_{1}^{x} -U_{0}^{y}\p_1 + \widetilde{U}_{1}^{y}=U_{1}^{x} + U_{1}^{y}=C$ we conclude
\begin{displaymath}
	\Delta \Q =  \epsilon \int_{\Omega} \V(\phi_{0})  d \psi d\phi_0 + \Og(\epsilon^2) \quad \text{where} \quad \V(\psi, \phi_{0})=C \E_{\phi_{0}}  .
\end{displaymath}

\end{document}